\documentclass[12pt,a4paper]{article}
\usepackage{jcappub}
\usepackage{tikz}
\usepackage{subcaption}

\makeatletter
\newcommand{\Spvek}[2][r]{%
  \gdef\@VORNE{1}
  \left[\hskip-\arraycolsep%
    \begin{array}{#1}\vekSp@lten{#2}\end{array}%
  \hskip-\arraycolsep\right]}

\def\vekSp@lten#1{\xvekSp@lten#1;vekL@stLine;}
\def\vekL@stLine{vekL@stLine}
\def\xvekSp@lten#1;{\def\temp{#1}%
  \ifx\temp\vekL@stLine
  \else
    \ifnum\@VORNE=1\gdef\@VORNE{0}
    \else\@arraycr\fi%
    #1%
    \expandafter\xvekSp@lten
  \fi}
\makeatother
\title{Non-linear Eulerian Hydrodynamics of Dark Energy: Riemann problem and Finite Volume Schemes}
\author[1]{Linda Blot}
\author[2]{Pier Stefano Corasaniti}
\author[1]{Fabian Schmidt}
\affiliation[1]{Max Planck Institute for Astrophysics, Karl-Schwarzschild str. 1, 85748 Garching, Germany}
\affiliation[2]{Laboratoire Univers et Th\'eories (LUTh), UMR 8102 CNRS, Observatoire de Paris, Universit\'e Paris Diderot, 5 Place   Jules Janssen, 92190 Meudon, France}
\emailAdd{lblot@mpa-garching.mpg.de}
\emailAdd{pier-stefano.corasaniti@obspm.fr}
\emailAdd{fabians@mpa-garching.mpg.de}
\abstract{Upcoming large-scale structure surveys can shed new light on the properties of dark energy. In particular, if dark energy is a dynamical component, it must have spatial perturbations. Their behaviour is regulated by the speed of sound parameter, which is currently unconstrained. In this work, we present the numerical methods that will allow to perform cosmological simulations of inhomogeneous dark energy scenarios where the speed of sound is small and non-vanishing. We treat the dark energy component as an effective fluid and build upon established numerical methods for hydrodynamics to construct a numerical solution of the effective continuity and Euler equations. In particular, we develop conservative finite volume schemes that rely on the solution of the Riemann problem, which we provide here in both exact and approximate forms for the case of a dark energy fluid.}
\begin{document}
\maketitle

\section{Introduction}
We have compelling evidence that the universe is currently undergoing an accelerated phase of expansion. In the standard cosmological model, this requires the existence of an exotic component, dubbed dark energy, that is characterized by negative pressure and dominates the present-day cosmic energy budget. A simple cosmological constant term ($\Lambda$) in Einstein's equation of General Relativity can play the role of dark energy since it behaves as a negative pressure component. Together with the Cold Dark Matter (CDM) hypothesis, the $\Lambda$CDM scenario currently reproduces all cosmological observations on the scales thus far available \cite{Planck2018,Boss2017,DESY3,HSC2020}. Nevertheless, the success of this model remains purely empirical as the physical origin of the cosmic dark components remains unknown. 

Observations indicate that the energy density associated with $\Lambda$ is $\rho_\Lambda\approx 10^{-47}$ GeV$^4$. It is the smallness of this value that has so far posed puzzling problems to any attempt of finding a satisfying theoretical explanation of $\Lambda$ (for a review see \cite{Carroll2001}). On the observational side, the $\Lambda$CDM model is showing its limitations in the form of tensions between cosmological parameter values recovered from different kinds of cosmological probes \cite{Verde2019,KiDS2021}. For these reasons, alternative hypotheses to the cosmological constant have been advanced. Beyond the cosmological constant, dark energy may result either from a modification of gravity on cosmic scales \cite{Tsujikawa2010,Clifton2012} or be the effect of a new degree of freedom not accounted for by the Standard Model of particles \cite{Copeland2006}. 

Given the lack of new fundamental principles which may guide the dark energy theoretical model building, there is hope nevertheless that the next generation of cosmological observations may reveal hints of the new physics underlying the dark energy phenomenon. Ongoing and upcoming large-scale structure surveys have the potential to constrain dark energy properties using established probes such as galaxy clustering, cosmic shear, and galaxy cluster number counts as well as newly proposed probes such as voids, counts in cells, and others. The usual approach is to model dark energy as a perfect fluid characterized by homogeneous pressure and density that are related through an equation of state parameter, $w$, the cosmological constant case corresponding to $w=-1$. However, if dark energy is dynamical, i.e. its energy density is not constant in time, general covariance imposes that it has to vary in space as well. We then have an additional macroscopic quantity that specifies the properties of the dark energy fluid, namely the speed of sound of dark energy perturbations, $c_s$. This suggests that the detection of the clustering of dark energy through observations of the cosmic structures may provide smoking gun evidence of the dynamical nature of the dark energy phenomenon.

Bounds on the dark energy equation of state and speed of sound can be translated into constraints on the microscopic properties of dark energy since these quantities can be related to terms in a Lagrangian effective field theory of dark energy (see e.g. \cite{Creminelli2009,Gleyzes2014}). For instance, quintessence models in which dark energy results from the dynamics of a minimally coupled scalar field are characterized by $c_s$ equal to the speed of light $c$ \cite{Caldwell1998}, while k-essence models in which dark energy is due to a scalar field with a non-canonical kinetic term have $c_s^2\approx 0$ \cite{ArmendarizPicon2001}. It is important to notice that a perfect fluid approach is not limited to the phenomenological study of dark energy. For instance models of unified dark matter and dark energy \cite{Bertacca2007} as well as alternative Dark Matter scenarios (see e.g. \cite{Khoury2014}) are described as dark fluids characterized by an equation of state and a speed of sound. For this reason throughout the text we may refer indistinctly to Dark Energy or dark fluids.

The clustering properties of Dark Energy can be tested through observations of the cosmic distribution of matter at different scales and times. On large scales, linear perturbation theory can be used to predict the effect of dark energy inhomogeneities on the evolution of matter density perturbations \cite{Erickson2001,BeanDore2004,Corasaniti2005}. However, since dark energy is a late-time phenomenon, the impact of dark energy perturbations on those scales is small and well within cosmic variance uncertainties, so that current data remains uninformative \cite{Corasaniti2005,dePutter2010}.

This may not be the case at small scales and late times where the collapse process is highly non-linear. For quintessence-like fluids, dark energy perturbations on sub-horizon scales have been damped by free-streaming, so that only the background expansion is affected. Thus, in these models the non-linear clustering of matter can be investigated using standard N-body simulations \cite{Alimi2010,Jennings2010,Courtin2011}. Instead, in models with $c_s<c$ perturbations may cluster at small scales so that it is necessary to include them in the simulation to capture the fully non-linear process of structure collapse in the presence of dark energy inhomogeneities. Previous studies looked at the fully clustered case $c_s^2=0$ in the context of the spherical collapse model \cite{Mota2004,Creminelli2010,Basse2011} or higher-order perturbation theory \cite{Sefusatti2011,Anselmi2011,Anselmi2014}. Recently the first cosmological simulation code that includes the effects of clustering dark energy has been developed, $k$-evolution \cite{Hassani2019,Hassani2020}. This is a relativistic N-body code that treats the dark energy component as a scalar field. While it has the advantage of including all the relevant relativistic effects, it cannot deal with the large dynamic range of typical high-resolution cosmological simulations since it employs a fixed grid that cannot be refined. Moreover, Ref.s \cite{Hassani2022a,Hassani2022b} reported instabilities in the scalar field solver in the context of cosmological simulations, providing strong motivation for an independent check of whether these instabilities are a physical property of clustering dark energy.

The work presented here is complementary to $k$-evolution in that it uses the Newtonian approximation but treats dark energy as a fluid and is meant to be implemented in fully non-linear hydrodynamics codes. In fact, we will show that the non-linear evolution of dark energy fluctuations is described by hyperbolic Euler equations which can be cast in the form of modified equations of hydrodynamics. Numerical schemes that are best suited to solve this type of equations are based on Finite Volume (FV) methods. Many of these schemes are already implemented in cosmological simulation codes such as ENZO \cite{BryanNorman1997}, FLASH \cite{Fryxell2000} or RAMSES \cite{Teyssier2002} to follow the cosmological collapse of baryonic gas. These codes use high-order variants of the Godunov scheme originally introduced in \cite{Godunov1959} and have been shown to provide accurate solutions to a variety of problems in gas dynamics. The accuracy of these schemes relies on their ability to capture discontinuities in the fluid flow and correctly predict the velocity of propagating waves. The starting point of these methods is the solution to the Riemann problem. Exact and approximate solutions have been derived for several systems including real gases \cite{ColellaGlaz1985}, inviscid flows of perfect gases \cite{GottliebGroth1988}, gases with generic equations of state \cite{MenikoffPlohr1989} and compressible liquids \cite{Ivings1998}. 

Here, we present a detailed study of the Riemann problem for cosmic dark fluid as a first step toward a cosmological study of the non-linear structure formation in inhomogeneous dark energy models. We derive an exact solution to the Riemann problem and construct several approximated solvers that in combination with Godunov-type schemes can efficiently solve the non-linear fluid equations for dark energy.

The paper is organized as follows: in Section~\ref{eulerequation} we present the Euler equations for a clustering dark energy fluid and cast them in the form that allows for an easier numerical implementation. We then present the solution of the Riemann problem for this fluid in Section~\ref{Riemann} and develop numerical Riemann solvers in Section~\ref{riemann_solvers}. Finally, in Section~\ref{upwind} we present working examples of finite volume conservative schemes that can be used to solve the Euler equations for clustering dark energy and test them using standard hydrodynamical test cases. We draw our conclusions in Section~\ref{conclu}.

\section{Euler Equations for Dark Energy Fluids}\label{eulerequation}

The equations describing the evolution of a dark energy (or any non-relativistic) fluid under the influence of Newtonian gravity in a Friedmann-Lemaitre-Robertson-Walker (FLRW) background have been derived in \cite{Sefusatti2011}, using conformal-Newtonian gauge. In Cartesian comoving coordinates these read as:
\begin{eqnarray}
&\frac{\partial\rho}{\partial\tau}&+3\mathcal{H}\left(\rho+\frac{p}{c^2}\right)+\vec{\nabla}\cdot\left[\left(\rho+\frac{p}{c^2}\right)\vec{v}\right]=0\label{cont1}\\
&\frac{\partial\vec{v}}{\partial\tau}&+\mathcal{H}\vec{v}+(\vec{v}\cdot\vec{\nabla})\vec{v}=-\frac{1}{\rho+\frac{p}{c^2}}\left(\vec{\nabla}p+\vec{v}\,\frac{\partial }{\partial\tau}\frac{p}{c^2}\right)-\vec{\nabla}\Phi,\label{eul1}
\end{eqnarray}
where $\rho$ and $p$ are the dark energy energy density and pressure respectively, $\vec{v}$ is the dark energy peculiar velocity with respect to the Hubble flow, $\tau$ is the conformal time defined by $d\tau = dt/a(t)$, with $a$ the scale factor, $\mathcal{H}=d\ln{a}/d\tau$ is the Hubble rate and $\Phi$ is the gravitational potential. The derivation of these equations assumes non-relativistic peculiar velocities ($v\ll c$), and scales much smaller than the horizon, which removes a number of relativistic contributions. 

Since we aim to use conservative numerical methods we want to cast the fluid equations in the form that is closest to the usual conservative form for baryonic fluids. To achieve this we can start by re-writing Eqs.~\eqref{cont1} and \eqref{eul1} as:
\begin{equation}
\frac{\partial}{\partial\tau}\left(\rho+\frac{p}{c^2}\right)+3\mathcal{H}\left(\rho+\frac{p}{c^2}\right)+\vec{\nabla}\cdot\left[\left(\rho+\frac{p}{c^2}\right)\vec{v}\right]=\frac{\partial p}{\partial\tau}
\end{equation}
\begin{multline}
\frac{\partial}{\partial\tau}\left[\left(\rho+\frac{p}{c^2}\right)\vec{v}\right]+4\mathcal{H}\left(\rho+\frac{p}{c^2}\right)\vec{v}+\vec{\nabla}\cdot\left[\left(\rho+\frac{p}{c^2}\right)\vec{v}\otimes\vec{v}\right]=\\
-\left(\rho+\frac{p}{c^2}\right)\vec{\nabla}\Phi-\vec{\nabla}p .
\end{multline}
Given the non-barotropic effective description of the dark energy fluid, it is necessary to split the evolution of the background density $\bar{\rho}$ and pressure $\bar{p}$ from that of the fluctuations $\delta\rho$ and $\delta p$ such that 
\begin{eqnarray}
\rho(\vec{x},\tau)&\equiv&\bar{\rho}(\tau)+\delta\rho(\vec{x},\tau)\\
p(\vec{x},\tau)&\equiv&\bar{p}(\tau)+\delta p(\vec{x},\tau),\label{dp}
\end{eqnarray}
where $\bar{p}(\tau)=w(\tau)\bar{\rho}(\tau)c^2$ and the evolution of the background density is governed by
\begin{equation}
\frac{\partial\bar{\rho}}{\partial\tau}=-3\mathcal{H}(1+w)\bar{\rho} .
\end{equation}
We then describe the dark energy pressure perturbations by relating them to dark energy density perturbations \emph{in the dark energy rest frame}, denoted by a subscript rf:
\begin{equation}
\delta p_{\rm rf}(\vec x,\tau) = c_s^2(\tau) \delta\rho_{\rm rf}(\vec x,\tau),
\end{equation}
where $c_s(\tau)$ is the effective sound speed, which in general can be time-dependent.

We can now introduce the new variable:
\begin{equation}
\Pi\equiv 1+w+\left(1+\frac{c_s^2}{c^2}\right)\delta,
\end{equation}
where $\delta=\delta\rho/\bar{\rho}$. We will restrict here to the case $\Pi>0$ because the solution of the Riemann problem becomes ill-defined for $\Pi\le0$ (see e.g. Eq.~(\ref{rightw}) and (\ref{leftw})), but there is also a physical reason for restricting to this case. In fact, one can also write $\Pi$ as $\Pi=(\rho_{\rm rf}+p_{\rm rf}/c^2)/\bar{\rho}_{\rm rf}$ and imposing that $(\rho+p/c^2)_{\rm rf} >0$ is equivalent to enforcing the null-energy condition. Notice that this translates to a floor in the physical dark energy underdensity:
\begin{equation}
    \delta>-\frac{1+w}{1+\frac{c_s^2}{c^2}},
\end{equation} 
meaning that the total dark energy density can never reach $0$ if $w<0$. Eq.~(\ref{cont1}) and Eq.~(\ref{eul1}) can finally be written in the form:
\begin{equation}
\frac{\partial\Pi}{\partial\tau}+\left(1+\frac{c_s^2}{c^2}\right)\vec{\nabla}\cdot\left(\Pi\vec{v}\right)=3\mathcal{H}\left(w-\frac{c_s^2}{c^2}\right)\left(\Pi-1-w\right)\label{cont3}
\end{equation}
\begin{equation}
\frac{\partial(\Pi\vec{v})}{\partial\tau}+\vec{\nabla}\cdot(\Pi\vec{v}\otimes\vec{v})+\frac{c_s^2}{\left(1+\frac{c_s^2}{c^2}\right)}\vec{\nabla}\Pi=\mathcal{H}(3w-1)\Pi\vec{v}-\Pi\vec{\nabla}\Phi,\label{eul3}
\end{equation}
which we call \textit{quasi-conservative} in analogy with the conservative form of the Euler equations for baryonic fluids. In fact, in the case of a vanishing equation of state $w$ and sound speed $c_s$, Eqs.~(\ref{cont3}) and~(\ref{eul3}) reduce to the standard conservation laws of mass and momentum. Here, we have intentionally written the friction terms due to the cosmic expansion on the right-hand side to distinguish the advection part of the Euler equations from non-advecting time-dependent source terms. We have also assumed that $w$ and $c_s$ are constant in time. Relaxing this assumption will introduce additional source terms. However, even if $c_s$ varies in time, this does not alter the form of the advection part of the Euler equations. Hence, the derivations presented hereafter remain valid also in this case. 

This system of equations gives a description of the gravitational dynamics of dark energy perturbations in the presence of other matter components provided a closure equation for the gravitational potential $\Phi$. This is given by the Poisson equation \citep[see e.g.][]{Sefusatti2011}:
\begin{equation}
    \nabla^2\Phi=4\pi G a^2 \left( \rho_{tot}+ 3 \frac{p_{tot}}{c^2} \right)
\end{equation}
where $\rho_{tot}$ and $p_{tot}$ are the total energy density and pressure of all gravitating components. 
Similar to what is done in the case of cosmological hydrodynamical simulations of dark matter and baryons, the idea of simulating the clustering of dark matter and dark energy is to use the N-body method to solve the dynamics of dark matter particles and the hydrodynamical description given by given by Eq.~(\ref{cont3}) and (\ref{eul3}) to solve that of dark energy fluctuations, where the same total potential, sourced by matter and dark energy fluctuations, is used for both components. These equations can be implemented for example in cosmological N-body/hydro codes with an Adaptive Mesh Refinement (AMR) of the simulation volume in which the energy density and pressure of dark energy can be added to that of the matter density field on the same AMR grid before numerically solving the Poisson equation.
\section{The Riemann Problem}\label{Riemann}

The advection parts of Eqs.~(\ref{cont3}) and~(\ref{eul3}) form a hyperbolic system of partial differential equations that can be solved numerically using finite volume methods. Here, we will focus on the Godunov method originally introduced in \citep{Godunov1959}. At the heart of the method, as we will see in detail in Section~\ref{upwind}, is the solution of the Riemann Problem (RP).

Let us briefly introduce here what is the Riemann Problem and how the solution is found. We can start by writing Eqs.~(\ref{cont3}) and~(\ref{eul3}) in a compact state-vector form in Cartesian coordinates:
\begin{equation}
\textbf{U}_\tau+\textbf{F}(\textbf{U})_x+\textbf{G}(\textbf{U})_y+\textbf{H}(\textbf{U})_z=\textbf{S}(\textbf{U}),\label{vec_eq}
\end{equation}
where the indices denote partial derivatives, $\textbf{U}$ is a state vector of unknowns, $\textbf{F}(\textbf{U})$, $\textbf{G}(\textbf{U})$ and $\textbf{H}(\textbf{U})$ are vectors of fluxes and $\textbf{S}(\textbf{U})$ a vector of sources. Let us focus on the advection part of the equations, which is simply given by imposing $\textbf{S}(\textbf{U})=0$. Physically, this corresponds to the case of a non-expanding background without gravity, which is what we consider in the following. Since in most applications this system is solved along each Cartesian direction separately, let us simply focus on the $x$ direction.

The Riemann problem is an initial value problem (IVP) of the advection equation
\begin{equation}
\textbf{U}_\tau+\textbf{F}(\textbf{U})_x=0,\label{IVP_Euler}
\end{equation}
with initial conditions
\begin{equation}
\textbf{U}(x,0)=
    \begin{cases}
    \textbf{U}_{\rm L} & \quad \text{if $x<0$}\\
    \textbf{U}_{\rm R} & \quad \text{if $x>0$}
    \end{cases} 
\end{equation}
In general, the solution of the RP for a $m\times m$ non-linear hyperbolic system consists of $m+1$ constant states separated by $m$ waves in the plane $x-\tau$. These can be either shocks, rarefaction or contact waves associated with the so called \textit{characteristic fields} of the advection equation Eq.~(\ref{IVP_Euler}) \citep[see e.g.][for a standard textbook presentation]{LeVeque2004,Toro2009}. The characteristics are curves in the plane $x-\tau$ along which Eq.~(\ref{IVP_Euler}) reduces to an ordinary differential equation. In order to determine these curves let us rewrite Eq.~(\ref{IVP_Euler}) as
\begin{equation}
\textbf{U}_\tau+\textbf{A}(\textbf{U})\textbf{U}_x=0,
\end{equation}
where $\textbf{A}(\textbf{U}) = \partial\textbf{A(U)}/\partial\textbf{U}$ is the Jacobian matrix of the flux. Finding the eigenvalues $\lambda_i$ and associated right eigenvectors $\textbf{R}_i$ of the Jacobian we can verify the nature of the waves by computing:
\begin{equation}
\vec{\nabla}_\textbf{U}\lambda_i\cdot\textbf{R}_i,
\end{equation}
where $\vec{\nabla}_\textbf{U}$ indicates the derivative with respect to the state variables $\textbf{U}$. If $\vec{\nabla}_\textbf{U}\lambda_i\cdot\textbf{R}_i\ne 0$ the characteristics are said to be \textit{genuinely non-linear} and the associated waves are either shocks or rarefaction waves, otherwise the wave is a contact discontinuity.

\begin{figure}
\centering
\begin{subfigure}[t]{0.3\textwidth}
\begin{tikzpicture}
  \draw[->] (0,0)--(4,0) node[right]{$x$};
  \draw[->] (0,0)--(0,4.0) node[above]{$\tau$};
  \draw[line width=1.5pt](0.0,0.0) -- (70:4cm) node[anchor=south west]{$s_T$};
  \draw[line width=0.5pt, color=gray](0.0,0.0) -- (65:4cm);
  \draw[line width=0.5pt, color=gray](0.0,0.0) -- (60:4cm);
  \draw[line width=0.5pt, color=gray](0.0,0.0) -- (55:4cm);
  \draw[line width=0.5pt, color=gray](0.0,0.0) -- (50:4cm);
  \draw[line width=0.5pt, color=gray](0.0,0.0) -- (45:4cm);
  \draw[line width=0.5pt, color=gray](0.0,0.0) -- (40:4cm);
  \draw[line width=1.5pt](0.0,0.0) -- (35:4cm) node[anchor=south west]{$s_H$};
  \node[align=center] at (2.5,1.0) {$\textbf{U}_{\rm R}$};
  \node[align=center] at (1.5,2.0) {$\textbf{U}_{fan}$};
  \node[align=center] at (0.5,2.7) {$\textbf{U}_{\rm L}$};
  \draw [-stealth](0.9,3.0) -- (0.7,4.0);
  \draw [-stealth](3.0,1.7) -- (4.0,1.8);
\end{tikzpicture}
\caption{Rarefaction wave: the left and right states are separated by the rarefaction fan, whose head and tail (thick lines) move with speed $s_H$ and $s_T$ respectively.}
\label{figRare}
\end{subfigure}
\hfill
\begin{subfigure}[t]{0.3\textwidth}
\centering
\begin{tikzpicture}
  \draw[->] (0,0)--(4,0) node[right]{$x$};
  \draw[->] (0,0)--(0,4) node[above]{$\tau$};
  \draw[line width=1.5pt](0,0)--(45:4cm) node[anchor=south west]{$s_s$};
  \node[align=center] at (3,1.0) {$\textbf{U}_{\rm R}$};
  \node[align=center](A) at (1.0,3.0) {$\textbf{U}_{\rm L}$};
  \draw [-stealth](0.5,2.0) -- (1.7,2.2);
  \draw [-stealth](2.0,0.5) -- (2.2,1.7);
\end{tikzpicture}
\caption{Shock wave: the left and right states are separated by a jump-discontinuity (thick line) that moves with speed $s_s$.}
\label{figShock}
\end{subfigure}
\hfill
\begin{subfigure}[t]{0.3\textwidth}
\centering
\begin{tikzpicture}
  \draw[->] (0,0)--(4,0) node[right]{$x$};
  \draw[->] (0,0)--(0,4) node[above]{$\tau$};
  \draw[dashed, line width=1.5pt](0,0)--(45:4cm) node[anchor=south west]{$s_c$};
  \node[align=center] at (3,1.0) {$\textbf{U}_{\rm R}$};
  \node[align=center](A) at (1.0,3.0) {$\textbf{U}_{\rm L}$};
  \draw [-stealth](0.5,1.0) -- (1.7,2.2);
  \draw [-stealth](1.0,0.5) -- (2.2,1.7);
\end{tikzpicture}
\caption{Contact wave: the left and right states are separated by a jump-discontinuity (thick dashed line) that moves with speed $s_c$.}
\label{figContact}
\end{subfigure}
\caption{Structure of the solution for the three types of waves. The directions of the characteristics on the two sides of the waves are indicated by the arrows.}
\end{figure}
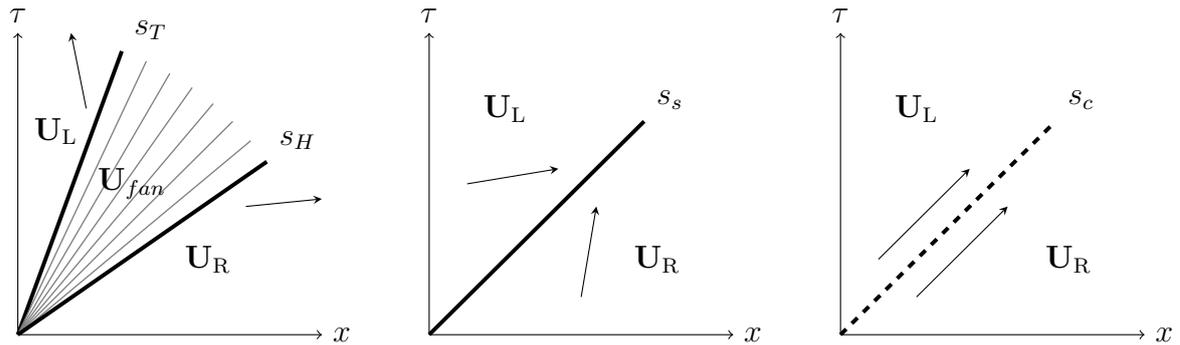

When $\lambda_i$ is associated to a rarefaction wave the states to the left and right of the wave $\textbf{U}_{\rm L}$ and $\textbf{U}_{\rm R}$ are connected through a smooth transition. The characteristics at the two sides of the wave diverge so that a new region emerges: the rarefaction fan. This region is enclosed between the head and the tail of the fan, whose speeds are given by $s_{H}=\lambda_i(\textbf{U}_{\rm R})$ and $s_{T}=\lambda_i(\textbf{U}_{\rm L})$ respectively. This is depicted schematically in Figure~\ref{figRare}.

The solution across rarefaction waves can be found using the generalized Riemann Invariants (see \cite{Jeffrey1976} for a detailed discussion). These are ordinary differential equations that relate the changes in the state variables to the respective components of the $i$-th eigenvector:
\begin{equation}
\frac{du_1}{r_1^i}=\frac{du_2}{r_2^i}=...=\frac{du_m}{r_m^i},\label{RI}
\end{equation}
where $u_i$ are the elements of the state vector $\textbf{U}$.

When the wave is a shock the characteristics at the two sides of the wave converge and the left and right states $\textbf{U}_{\rm L}$ and $\textbf{U}_{\rm R}$ are connected through a single jump-discontinuity that moves with speed $s_s$, as shown in Figure~\ref{figShock}. In this case we can make use of the Rankine-Hugoniot conditions, that relate the state variables on the two sides of the discontinuity:
\begin{equation}
\Delta\textbf{F}=s\Delta\textbf{U}.\label{RKHU}
\end{equation}

In case the wave is a contact discontinuity the states to the left and right of the wave are connected through a single jump-discontinuity, as shown in Figure~\ref{figContact}. The characteristics on the two sides run parallel to the wave, which moves with speed $s_c=\lambda_i(\textbf{U}_{\rm L})=\lambda_i(\textbf{U}_{\rm R})$. We can use either the Riemann Invariants or the Rankine-Hugoniot conditions to find the solution across contact waves.

\subsection{Structure of the Solution in 1D}
Let us now focus on the specific case of the Dark Energy fluid and find the analytical solution of the Riemann problem. In Section~\ref{exact_riemann} we will provide the full solution, however it is useful to first derive a set of equations that relate state variables across different wave patterns. Hereafter, we will extensively follow Toro's textbook \cite{Toro2009} to which we refer interested readers for a detailed presentation of numerical methods in fluid dynamics.

The vectors in Eq.~\eqref{vec_eq} are given by:
\begin{equation}
\textbf{U}=\Spvek[c]{u_1;u_2},\, \textbf{F}=\Spvek[c]{(1+\frac{c_s^2}{c^2})u_2;\frac{u_2^2}{u_1}+\frac{c_s^2}{1+\frac{c_s^2}{c^2}}u_1},\,
\textbf{S}=\Spvek[c]{3\mathcal{H}\left(w-\frac{c_s^2}{c^2}\right)\left(u_1-1-w\right);\mathcal{H}(3w-1)u_2-u_1\frac{\partial \Phi}{\partial x}},
\end{equation}
where $u_1=\Pi$ and $u_2=\Pi v$ are the conservative state variables. Since this is a $2\times 2$ system of equations the solution consists of 3 states separated by 2 waves.
The Jacobian matrix reads:
\begin{equation}\label{jacobian1Dcase}
\textbf{A}(\textbf{U})\equiv\frac{\partial\textbf{F}}{\partial\textbf{U}}=
\left[\begin{array}{cc}
0 & 1+\frac{c_s^2}{c^2} \\
 \frac{c_s^2}{1+\frac{c_s^2}{c^2}}-v^2 & 2v
\end{array}\right].
\end{equation}
and its eigenvalues are given by 
\begin{equation}
\lambda_\pm=v\pm c_s\sqrt{1-\frac{v^2}{c^2}}\label{lambda}.
\end{equation}
These are real and distinct if and only if $c_s> 0$. In such a case we have $\lambda_-(v)<\lambda_+(v)$ and the system is \textit{strictly hyperbolic}. Moreover, since we are considering Dark Energy models for which peculiar velocities are non-relativistic ($v\ll c$) the wave structure of the system approximately reproduces that of an isothermal fluid with constant sound speed, for which $\lambda_{\pm}=v\pm c_s$. It is also worth noticing that the condition of hyperbolicity requires a limiting speed for the dark fluid velocity $v\le c$, which is direct consequence of including terms $p/c^2$ in Eq.~(\ref{cont1}) and (\ref{eul1}). The reader may wonder whether in the presence of gravitational interactions the speed of dark energy fluctuations may numerically become ultra-relativistic, thus leading to loss of hyperbolicity of the system. We believe this not to be the case, since these dark fluids are cold, with negligibly small velocities, and cosmological simulations generally have gravitational potentials $\Phi \ll 1$.

The right eigenvectors associated with $\lambda_\pm$ are given by
\begin{equation}
\textbf{R}_+=\Spvek[c]{1+\frac{c_s^2}{c^2};v+c_s\sqrt{1-\frac{v^2}{c^2}}},\, \textbf{R}_-=\Spvek[c]{1+\frac{c_s^2}{c^2};v-c_s\sqrt{1-\frac{v^2}{c^2}}},\label{eigenvec}
\end{equation}
while the left eigenvectors read as
\begin{eqnarray}
\textbf{L}_+&=&\left[\frac{1}{2\left(1+\frac{c_s^2}{c^2}\right)}\left(1-\frac{v}{c_s\sqrt{1-\frac{v^2}{c^2}}}\right),\,\,\,\,\,\,\,\frac{1}{2c_s\sqrt{1-\frac{v^2}{c^2}}}\right],\nonumber\\ &&\\
\textbf{L}_-&=&\left[\frac{1}{2\left(1+\frac{c_s^2}{c^2}\right)}\left(1+\frac{v}{c_s\sqrt{1-\frac{v^2}{c^2}}}\right),\,\,\,\,\,\,\,\frac{-1}{2c_s\sqrt{1-\frac{v^2}{c^2}}}\right]\nonumber
\end{eqnarray}
and satisfy the relation $\textbf{L}_\pm\cdot\textbf{R}_\pm=\delta_{\pm\pm}$, where $\delta_{\pm\pm}$ is the Kronecker symbol.

Using Eqs.~(\ref{lambda}) and (\ref{eigenvec}) we obtain
\begin{equation}
\vec{\nabla}_\textbf{U}\lambda_\pm\cdot\textbf{R}_\pm=\pm\frac{c_s}{\Pi\sqrt{1-\frac{v^2}{c^2}}}\left[1+\frac{v^2}{c^2}\left(1+\frac{c_s^2}{c^2}\right)\right],
\end{equation}
hence the characteristic fields are genuinely non-linear if and only if $c_s> 0$. In this case, the waves associated with $\lambda_\pm$ are either rarefaction or shock waves.

\begin{figure}
\centering
\begin{tikzpicture}
  \draw[->] (-4,0)--(4,0) node[right]{$x$};
  \draw[->] (0,0)--(0,5) node[above]{$\tau$};
  \draw[line width=1.5pt](0,0)--(2.5,4) node[anchor=south west]{$\lambda_+$};
  \draw[line width=1.5pt](0,0)--(-2.5,4) node[anchor=south east]{$\lambda_-$};
  \node[align=center] at (0,-0.3) {$x_0$};
  \node[align=center] at (-3,1.5) {$\textbf{U}_{\rm L}$};
  \node[align=center] at (3,1.5) {$\textbf{U}_{\rm R}$};
  \node[align=center](A) at (-0.5,2.5) {$\textbf{U}^{*}$};
\end{tikzpicture}
\caption{Structure of the solution of the Riemann problem for a dark energy fluid in 1D: two waves $\lambda_{\pm}$ (thick lines) generate from the discontinuity at $x_0$ and a new state $\textbf{U}_*$ emerges in between them.}
\label{fig1}
\end{figure}
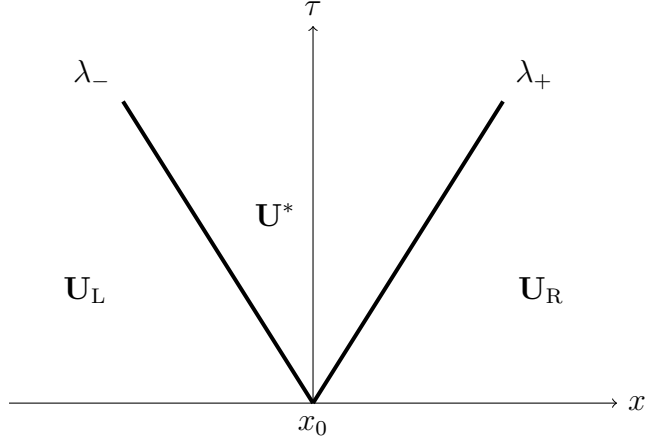

In Fig.~\ref{fig1} we sketch the structure of the solution to the Riemann problem. Given the data to the left $\textbf{U}_{\rm L}$ and the right $\textbf{U}_{\rm R}$ of the initial discontinuity at $x=x_0$, we want to determine the state $\textbf{U}^*$ in the region enclosed between the left and right waves and derive the solution of the Riemann problem $\textbf{U}(x,\tau)$. Once the nature of the left and right waves is known we can determine the value of $\textbf{U}^*$ as a function of the left or right state using the Generalized Riemann Invariants or the Rankine-Hugoniot conditions.

Notice that Eq.~(\ref{IVP_Euler}) can be rewritten also in terms of the primitive state variables $\textbf{W}=(\Pi,v)$:
\begin{equation}
\textbf{W}_\tau+\tilde{\textbf{A}}(\textbf{W})\textbf{W}_x=0,\label{advprimiv}
\end{equation}
where 
\begin{equation}
\tilde{\textbf{A}}(\textbf{W})=
\left[\begin{array}{cc}
\left(1+\frac{c_s^2}{c^2}\right)v & \left(1+\frac{c_s^2}{c^2}\right)\Pi \\
 \frac{c_s^2}{\Pi}\left(\frac{1}{1+\frac{c_s^2}{c^2}}-\frac{v^2}{c^2}\right) & \left(1-\frac{c_s^2}{c^2}\right)v
\end{array}\right].
\end{equation}
Since conservative and primitive variables are related by a linear transformation $d\textbf{U}=\Lambda \, d\textbf{W}$ with 
\begin{equation}
\Lambda=
\left[\begin{array}{cc}
1 & 0 \\
v & \Pi
\end{array}\right] \quad \quad
\Lambda^{-1}=
\left[\begin{array}{cc}
1 & 0 \\
-\frac{v}{\Pi} & \frac{1}{\Pi}
\end{array}\right]
\end{equation}
it is easy to show that $\textbf{A}(\textbf{U})=\Lambda\,\tilde{\textbf{A}}(\textbf{W})\Lambda^{-1}$. This implies that the Jacobian matrix, whether written in terms of primitive variables or conservative ones, has identical eigenvalues, while the corresponding eigenvectors are related by a linear transformation. Therefore, the Riemann problem admits the same wave structure in the two formulations.

In this case, the right eigenvectors are given by
\begin{equation}
\tilde{\textbf{R}}_+=\Spvek[c]{\left(1+\frac{c_s^2}{c^2}\right)\Pi;-\frac{c_s^2}{c^2} v+c_s\sqrt{1-\frac{v^2}{c^2}}},\, \tilde{\textbf{R}}_-=\Spvek[c]{\left(1+\frac{c_s^2}{c^2}\right)\Pi;-\frac{c_s^2}{c^2} v-c_s\sqrt{1-\frac{v^2}{c^2}}},\label{righteigenvec}
\end{equation}
while the left eigenvectors read as
\begin{eqnarray}
\tilde{\textbf{L}}_+&=&\left[\frac{1}{2\left(1+\frac{c_s^2}{c^2}\right)\Pi}\left(1+\frac{\frac{c_s^2}{c^2} v}{c_s\sqrt{1-\frac{v^2}{c^2}}}\right),\,\,\,\,\,\,\,\frac{1}{2c_s\sqrt{1-\frac{v^2}{c^2}}}\right],\nonumber\\ &&\label{lefteigenvec}\\
\tilde{\textbf{L}}_-&=&\left[\frac{1}{2\left(1+\frac{c_s^2}{c^2}\right)\Pi}\left(1-\frac{\frac{c_s^2}{c^2} v}{c_s\sqrt{1-\frac{v^2}{c^2}}}\right),\,\,\,\,\,\,\,\frac{-1}{2c_s\sqrt{1-\frac{v^2}{c^2}}}\right]\nonumber
\end{eqnarray}

In Appendix~\ref{app_rp1d} we derive the relations between the state variables across different wave patterns.
\subsection{Structure of the Solution in 3D}
In the previous section we discussed the Riemann problem in 1D. As we will show here this provides the basis to solve the Riemann problem in 3D. In fact, let us consider Eqs.~(\ref{cont3}) and (\ref{eul3}) without source terms in Cartesian coordinates. In terms of the conservative variable these can be written in a state-vector form:
\begin{equation}
\textbf{U}+\textbf{A}(\textbf{U})\textbf{U}_x+\textbf{B}(\textbf{U})\textbf{U}_y+\textbf{C}(\textbf{U})\textbf{U}_z=0,
\end{equation}
with $\textbf{U}=(\Pi,\Pi v_{x},\Pi v_{y},\Pi v_{z})$ where $v_x,v_y$ and $v_z$ are the components of the peculiar velocity vector along $x$, $y$ and $z$ respectively, and
\begin{equation}
\textbf{A}(\textbf{U})=
\left[\begin{array}{cccc}
0 & 1+\frac{c_s^2}{c^2} & 0 & 0 \\
 \frac{c_s^2}{1+\frac{c_s^2}{c^2}}-v_x^2 & 2v_x & 0 & 0\\
 -v_x v_y & v_y & v_x & 0 \\
 -v_x v_z & v_z & 0 & v_x 
\end{array}\right],\,
\textbf{B}(\textbf{U})=
\left[\begin{array}{cccc}
0 & 0 & 1+\frac{c_s^2}{c^2} & 0 \\
 -v_x v_y & v_y & v_x  & 0 \\
 \frac{c_s^2}{1+\frac{c_s^2}{c^2}}-v_y^2 & 0 & 2v_y & 0\\
 -v_y v_z & 0 & v_z & v_y 
\end{array}\right],\,\nonumber
\end{equation}
\begin{equation}
\textbf{C}(\textbf{U})=
\left[\begin{array}{cccc}
0 & 0 & 0 & 1+\frac{c_s^2}{c^2} \\
 -v_x v_z & v_z & 0 & v_x  \\
 -v_y v_z & 0 & v_z & v_y  \\
 \frac{c_s^2}{1+\frac{c_s^2}{c^2}}-v_z^2 & 0 & 0 & 2v_z
\end{array}\right].\nonumber
\end{equation}
The eigenvalues of these matrices are:
\begin{align*}
&\lambda_{1}=v-c_s\sqrt{1-\frac{v^2}{c^2}}\\
&\lambda_{2,3}=v\\
&\lambda_{4}=v+c_s\sqrt{1-\frac{v^2}{c^2}}
\end{align*} with $v=v_x,v_y$ and $v_z$ for $\textbf{A}$, $\textbf{B}$ and $\textbf{C}$ respectively. Hence, the 3D case can be seen as the composition of three advections with the same structure of the Riemann problem.

Let us consider the advection equations along the x-direction. The eigenvalues $\lambda_{1,4}$ are real and distinct and associated to the genuinely non-linear characteristic fields that correspond to shock and rarefaction waves, while $\lambda_{2,3}$ are linearly degenerate and describe a contact discontinuity. Moreover, the eigenvalues $\lambda_{1,4}$ are identical to the 1D case, so that the solution for $\Pi$ and $v_x$ will be the same as the one derived in Appendix~\ref{app_rp1d}. The right eigenvectors read as
\begin{align}
\textbf{R}_1=\Spvek[c]{1+\frac{c_s^2}{c^2};v_x-c_s\sqrt{1-\frac{v_x^2}{c^2}};\left(1+\frac{v_x c_s}{c^2\sqrt{1-\frac{v_x^2}{c^2}}}\right)v_y;\left(1+\frac{v_x c_s}{c^2\sqrt{1-\frac{v_x^2}{c^2}}}\right)v_z},\,&
\textbf{R}_4=\Spvek[c]{1+\frac{c_s^2}{c^2};v_x+c_s\sqrt{1-\frac{v_x^2}{c^2}};\left(1-\frac{v_x c_s}{c^2\sqrt{1-\frac{v_x^2}{c^2}}}\right)v_y;\left(1-\frac{v_x c_s}{c^2\sqrt{1-\frac{v_x^2}{c^2}}}\right)v_z},\,\nonumber\\
\textbf{R}_2=\Spvek[c]{0;0;1;0},\,& \textbf{R}_3=\Spvek[c]{0;0;0;1},\, \label{eigenvecA}
\end{align}
As can be seen from the eigenvectors $\textbf{R}_2$ and $\textbf{R}_3$ only the transverse velocity components change across the contact wave. The wave structure is depicted in Figure~\ref{fig:3Dwave}. In Appendix~\ref{app_rp3d} we derive the expressions for the transverse velocities across different wave patterns.

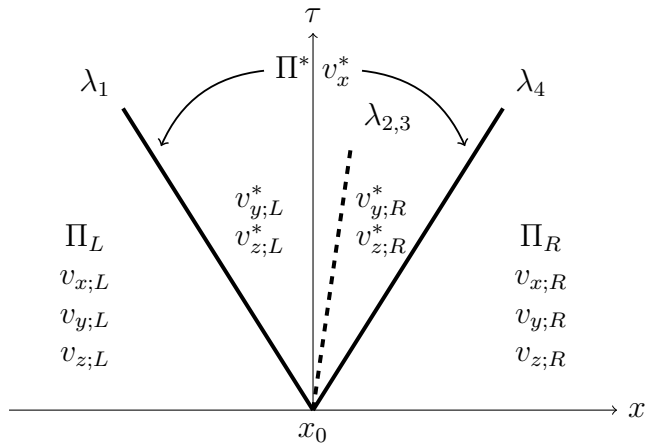
\begin{figure}
\centering
\begin{tikzpicture}
  \draw[->] (-4,0)--(4,0) node[right]{$x$};
  \draw[->] (0,0)--(0,5) node[above]{$\tau$};
  \draw[line width=1.5pt](0,0)--(2.5,4) node[anchor=south west]{$\lambda_4$};
  \draw[line width=1.5pt,dashed](0,0)--(0.5,3.5) node[anchor=south west]{$\lambda_{2,3}$};
  \draw[line width=1.5pt](0,0)--(-2.5,4) node[anchor=south east]{$\lambda_1$};
  \node[align=center] at (0,-0.3) {$x_0$};
  \node[align=center] at (-3,1.5) {$\Pi_L$ \\ $v_{x;L}$ \\ $v_{y;L}$ \\ $v_{z;L}$};
  \node[align=center] at (3,1.5) {$\Pi_R$ \\ $v_{x;R}$ \\ $v_{y;R}$ \\ $v_{z;R}$};
  \node[align=center](A) at (0,4.5) {$\Pi^*$ $v_x^*$};
  \node[align=center] at (-0.7,2.5) {$v_{y;L}^*$ \\ $v_{z;L}^*$};
  \node[align=center] at (0.9,2.5) {$v_{y;R}^*$ \\ $v_{z;R}^*$};
  \draw[->,line width=0.7pt] (A.west) to[bend right] (-2,3.5);
  \draw[->,line width=0.7pt] (A.east) to[bend left] (2,3.5);
\end{tikzpicture}
\caption{Structure of the solution of the 3D Riemann problem along the $x$ dimension. $\lambda_{1,4}$ are genuinely non-linear fields and can be either shock or rarefaction waves (thick lines). All state variables change across these waves. $\lambda_{2,3}$ are linearly degenerate fields and are associated to a contact discontinuity (thick dashed line). Only the transverse velocities $v_y$ and $v_z$ change across this wave.}
\label{fig:3Dwave}
\end{figure}

\section{Riemann Solvers}\label{riemann_solvers}
\subsection{Exact Riemann Solver}\label{exact_riemann}
Having derived relations between primitive state variables $\textbf{W}_{\rm L}$, $\textbf{W}_{\rm R}$ and $\textbf{W}^*$ across the different wave patterns, we can derive an exact solution of the RP for dark fluids as given by Eq.~(\ref{IVP_Euler}). Since the solution for the transverse velocities is trivial we will focus here on the 1D case.

The first step consists in determining the value of $\Pi^*$ in the star region of Fig.~\ref{fig1}. We have expressed the velocity $v^*$ in terms of $\Pi^*$ and the known states $\textbf{W}_{\rm L}$, $\textbf{W}_{\rm R}$ for all possible wave patterns. Thus, equating these relations for the left and right waves, $\Pi^*$ is the root of the algebraic equation
\begin{equation}
f_{\rm L}(\Pi^*,\textbf{W}_{\rm L})-f_{\rm R}(\Pi^*,\textbf{W}_{\rm R})=0,\label{exactriem}
\end{equation}
where 
\begin{equation}
f_{\rm L}(\Pi,\textbf{W}_{\rm L})=
    \begin{cases}
    f_{\rm Ls}(\Pi,\textbf{W}_{\rm L}) \quad \text{if $\Pi>\Pi_{\rm L}$ (shock)}\\
    f_{\rm Lr}(\Pi,\textbf{W}_{\rm L}) \quad \text{if $\Pi\le \Pi_{\rm L}$ (rarefaction)}
    \end{cases} 
\end{equation}
with $f_{\rm Ls}(\Pi,\textbf{W}_{\rm L})$ and $f_{\rm Lr}(\Pi,\textbf{W}_{\rm L})$ given by Eq.~(\ref{left_shock_rel}) and Eq.~(\ref{flr}) respectively, while
\begin{equation}
f_{\rm R}(\Pi,\textbf{W}_{\rm R})=
    \begin{cases}
    f_{\rm Rs}(\Pi,\textbf{W}_{\rm R}) \quad \text{if $\Pi>\Pi_{\rm R}$ (shock)}\\
    f_{\rm Rr}(\Pi,\textbf{W}_{\rm R}) \quad \text{if $\Pi\le \Pi_{\rm R}$ (rarefaction)}
    \end{cases} 
\end{equation}
with $f_{\rm Rs}(\Pi,\textbf{W}_{\rm R})$ and $f_{\rm Rr}(\Pi,\textbf{W}_{\rm R})$ given by Eq.~(\ref{right_shock_rel}) and Eq.~(\ref{frr}) respectively. Eq.~(\ref{exactriem}) can be solved for $\Pi^*$ to the desired level of accuracy using standard numerical root-finder schemes. Solving for $\Pi^{*}$ completely determines the nature of the waves. The value of $v^*$ can then be computed from one of the functions $f_{\rm L}(\Pi^*,\textbf{W}_{\rm L})$ or $f_{\rm R}(\Pi^*,\textbf{W}_{\rm R})$. Once the state $\textbf{W}^*$ is known, the solution of the Riemann problem across the entire spatial interval can be sampled at any given time as follows:

\begin{itemize}
\item in the region to the left of the discontinuity, corresponding to $x-x_0<v^* \tau$:
    \begin{itemize}
    \item if $\Pi_{\rm L}>\Pi^*$ the left wave is a rarefaction, then
    \begin{equation}
    \textbf{W}(x,\tau)=
     \begin{cases}
     (\Pi_{\rm L},v_{\rm L}) \quad \text{if $x-x_0<s_{\rm HLr}\tau$}\\
     (\Pi_{\rm Lfan},v_{\rm Lfan})\quad \text{if $s_{\rm HLr}\tau<x-x_0<s_{\rm TLr}\tau$}\\
      (\Pi^*,v^*)\quad \text{if $x-x_0>s_{\rm TLr}\tau$}\\
     \end{cases}
    \end{equation}
where $v_{\rm Lfan}$ is given by solving Eq.~(\ref{leftvfan}) and $\Pi_{\rm Lfan}$ is given by Eq.~(\ref{leftdfan}).
    \item if $\Pi_{\rm L}<\Pi^*$ the left wave is a shock, then
    \begin{equation}
    \textbf{W}(x,\tau)=
     \begin{cases}
     (\Pi_{\rm L},v_{\rm L}) \quad \text{if $x-x_0<s_{\rm Ls}\tau$}\\
      (\Pi^*,v^*)\quad \text{if $x-x_0>s_{\rm Ls}\tau$}\\
     \end{cases}
    \end{equation}
where $s_{\rm Ls}$ is given by Eq.~(\ref{leftvshock}).
    \end{itemize}
\item in the region to the right of the discontinuity corresponding to $x-x_0>v^* \tau$:
\begin{itemize}
    \item if $\Pi_{\rm R}>\Pi^*$ the right wave is a rarefaction, then
    \begin{equation}
    \textbf{W}(x,\tau)=
     \begin{cases}
     (\Pi^*,v^*) \quad \text{if $x-x_0<s_{\rm TRr}\tau$}\\
     (\Pi_{\rm Rfan},v_{\rm Rfan})\quad \text{if $s_{\rm TRr}\tau<x-x_0<s_{\rm HRr}\tau$}\\
      (\Pi_{\rm R},v_{\rm R})\quad \text{if $x-x_0>s_{\rm HRr}\tau$}\\
     \end{cases}
    \end{equation}
where $v_{\rm Rfan}$ is given by solving Eq.~(\ref{rightvfan}) and $\Pi_{\rm Rfan}$ is given by Eq.~(\ref{rightdfan}).
    \item if $\Pi_{\rm R}<\Pi^*$ the right wave is a shock, then 
    \begin{equation}
    \textbf{W}(x,\tau)=
     \begin{cases}
     (\Pi^*,v^*) \quad \text{if $x-x_0<s_{\rm Rs}\tau$}\\
      (\Pi_{\rm R},v_{\rm R})\quad \text{if $x-x_0>s_{\rm Rs}\tau$}\\
     \end{cases}
    \end{equation}
where $s_{\rm R}s$ is given by Eq.~(\ref{rightvshock}).
    \end{itemize}
\end{itemize}

\begin{table}[t]
\centering
\begin{tabular}{|c|c|c|c|c|}
\hline
Test & $\Pi_{\rm L}$ & $v_{\rm L}$ & $\Pi_{\rm R}$  & $v_{\rm R}$  \\
\hline
 $1$ &  $1.0$ & $0.0$ & $0.125$ & $0.0$ \\
\hline
 $2$ &  $1.0$ & $-0.0003$ & $1.0$ & $0.0003$ \\
\hline
 $3$ &  $0.3$ & $0.0001$ & $0.3$ & $-0.0001$ \\
\hline
\end{tabular}
\caption{Initial data for three test problems with exact solution.}\label{tab1}
\end{table}

We construct an exact Riemann solver as in \cite{Toro2009}. We first solve Eq.~(\ref{exactriem}) using the Newton-Raphson method. The value of $\Pi^*$ is obtained to a given level of accuracy $\epsilon$ through an iteration procedure
\begin{equation}
\Pi^{*}_{(i)}=\Pi^{*}_{(i-1)}-\frac{f[\Pi^{*}_{(i-1)}]}{f'[\Pi^{*}_{(i-1)}]},
\end{equation}
where $f(\Pi^{*})=f_{\rm L}(\Pi^*,\textbf{W}_{\rm L})-f_{\rm R}(\Pi^*,\textbf{W}_{\rm R})$ and $f'(\Pi^{*})=df/d\Pi|_{*}$ (the derivatives can be computed analytically). The iteration continues until the desired accuracy is reached
\begin{equation}
\frac{|\Pi^{*}_{(i)}-\Pi^{*}_{(i-1)}|}{|\Pi^{*}_{(i)}+\Pi^{*}_{(i-1)}|}<\frac{\epsilon}{2}.
\end{equation}
An initial guess value is necessary to start the iteration. Thus, the computation of the solution requires several iterations if the initial guess is too far off the solution. To address this point we use an adaptive scheme to optimize the initial guess using approximate Riemann solvers described in Section~\ref{appriemann}. 

\begin{figure}[ht]
\centering
\begin{tabular}{cc}
\includegraphics[width=5.5cm]{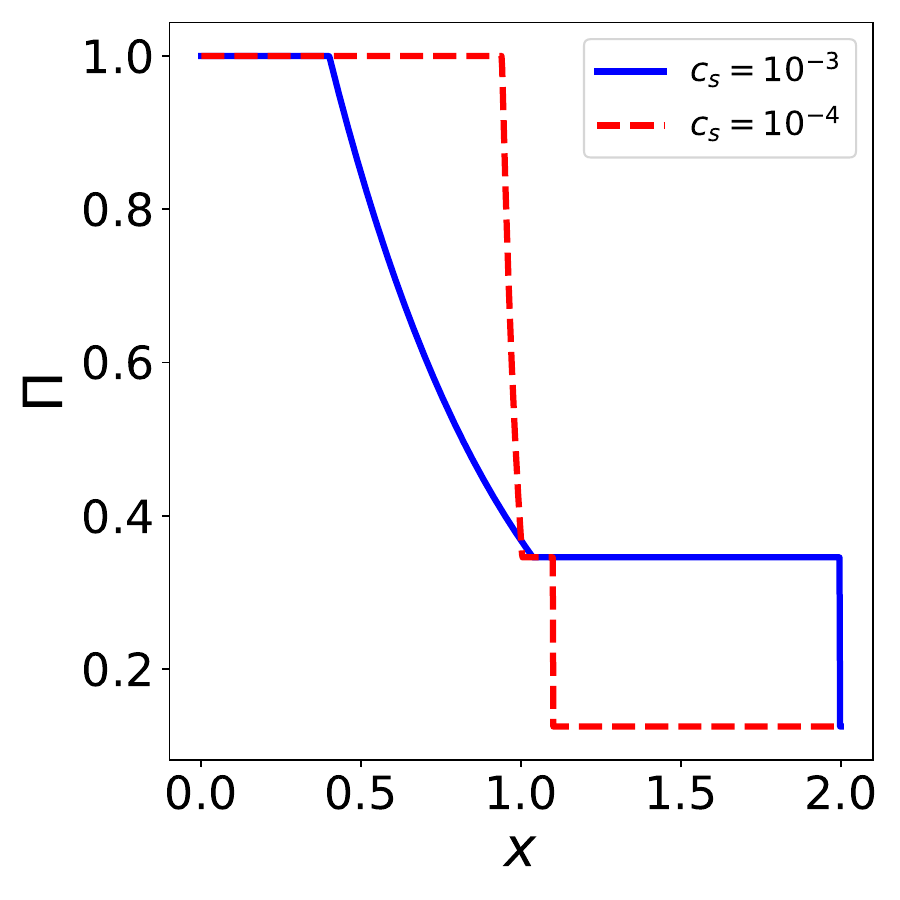} &
\includegraphics[width=5.5cm]{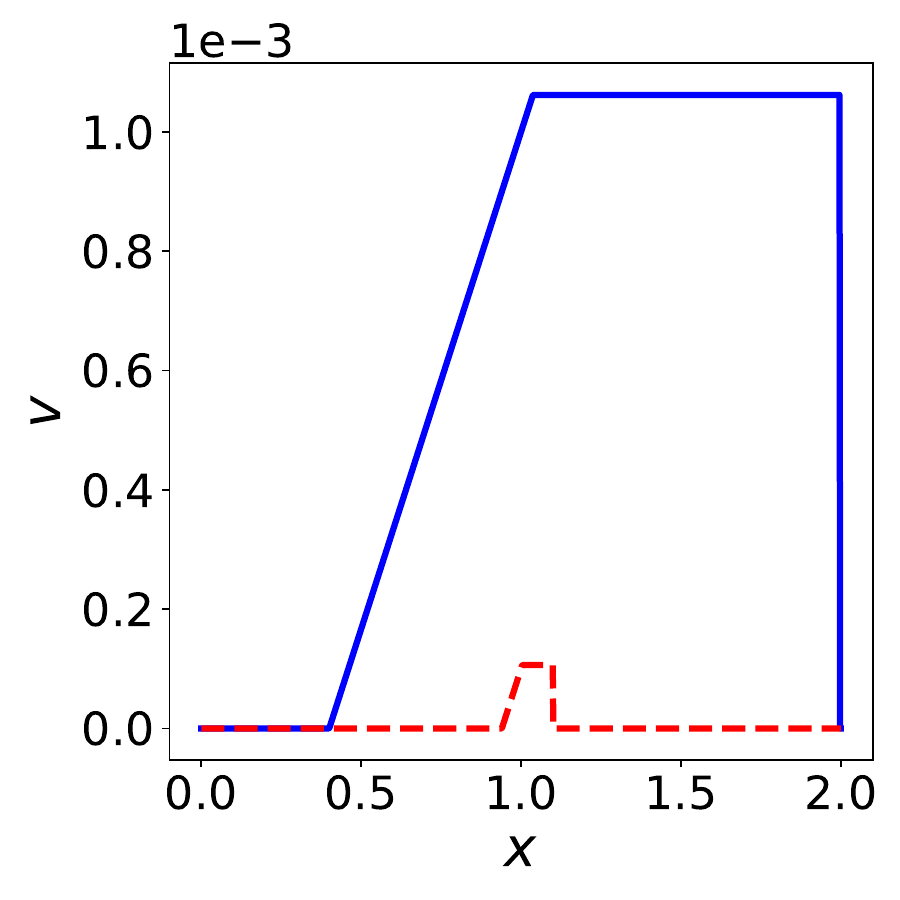} \\
\end{tabular}
\caption{Exact solution of Test 1 for $\Pi$ (left panel) and $v$ (right panel) at time $\tau=600$ for $c_s=10^{-3}$ (blue solid line) and $10^{-4}$ (red dashed line) respectively.\linebreak}\label{fig2}
\centering
\begin{tabular}{cc}
\includegraphics[width=5.5cm]{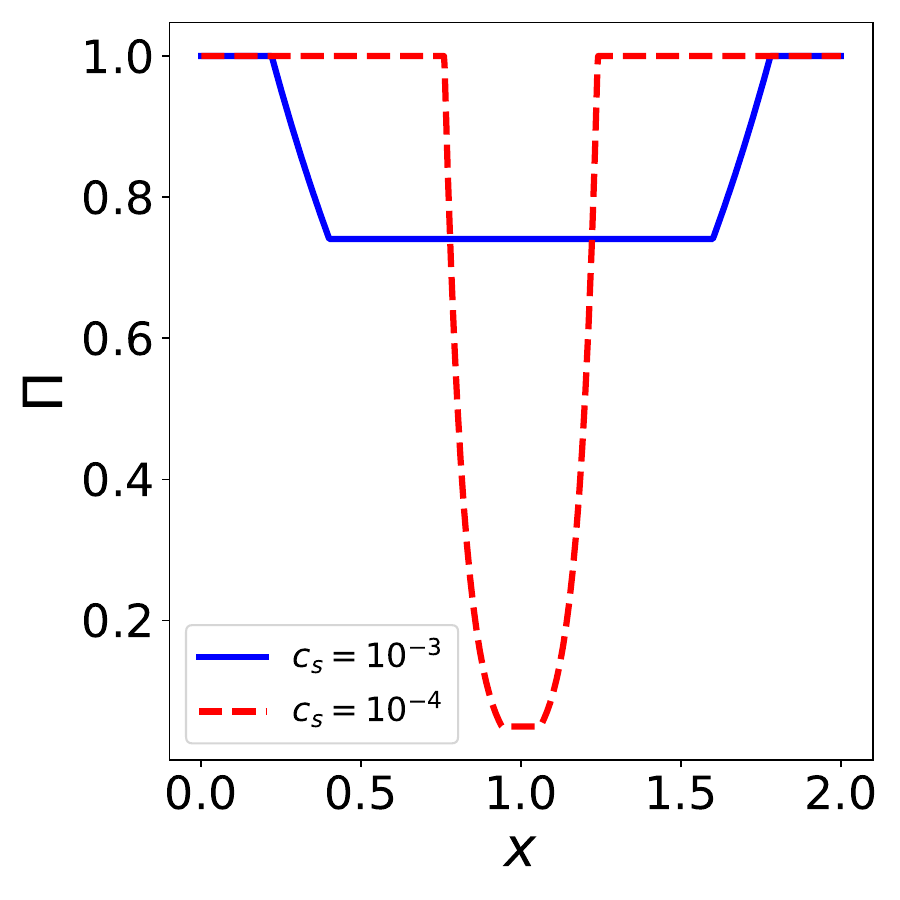}&
\includegraphics[width=5.5cm]{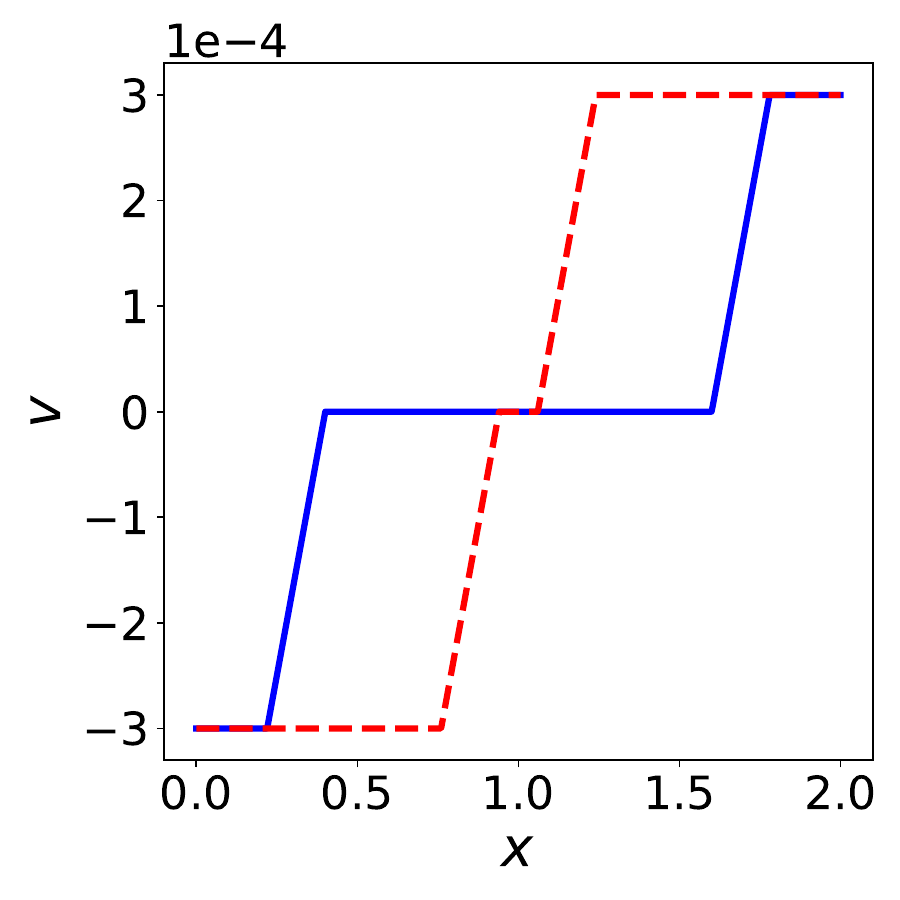} \\
\end{tabular}
\caption{As in Fig.~\ref{fig2} for Test 2.\linebreak}\label{fig3}
\centering
\begin{tabular}{cc}
\includegraphics[width=5.5cm]{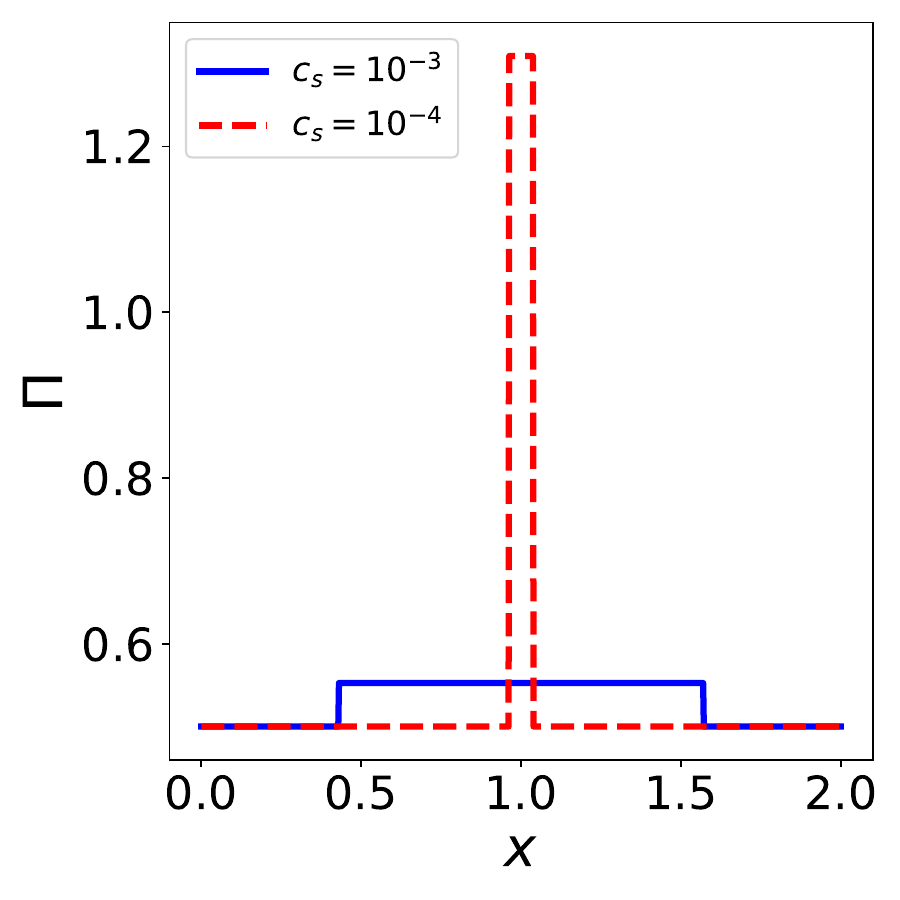}&
\includegraphics[width=5.5cm]{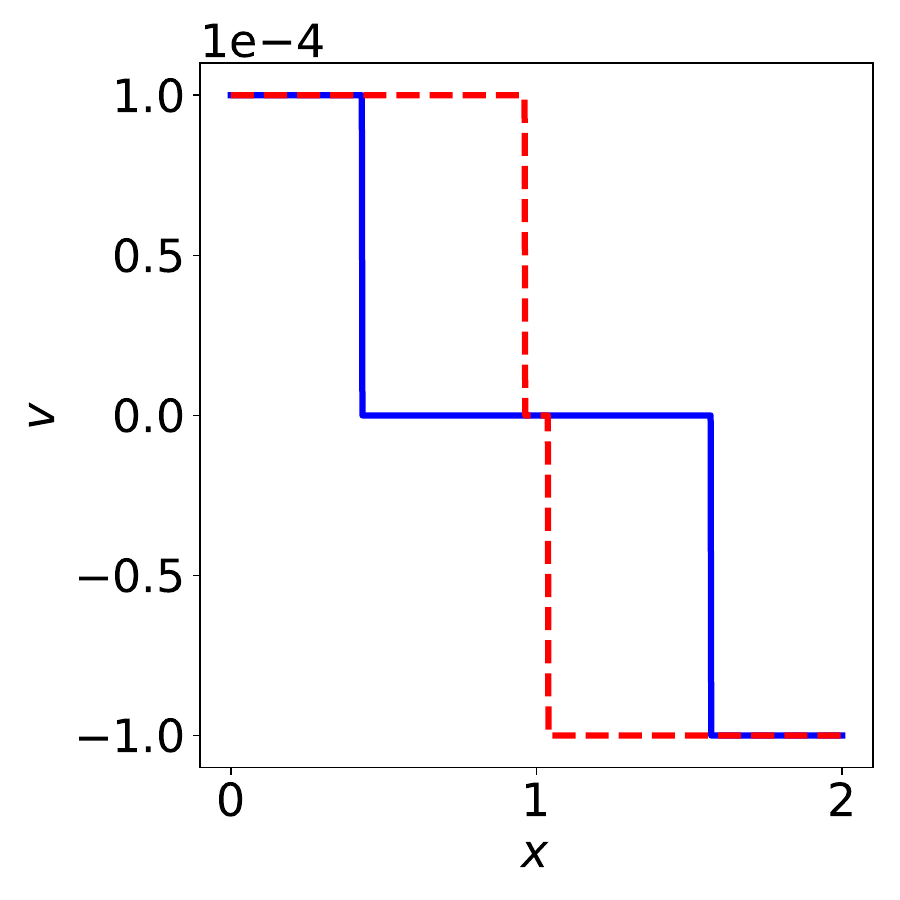} \\
\end{tabular}
\caption{As in Fig.~\ref{fig2} for Test 3.}\label{fig4}
\end{figure}

In Table~\ref{tab1} we quote the initial data for three standard test cases of the RP in arbitrary units. Notice that the equation of state of the fluid only enters via the source terms in the equations for $\Pi$ and $u$ (Equation~\ref{cont3}), and thus does not appear in the RP test cases. Test 1 is the standard Sod test \cite{Sod1978}, whose solution consists of a left rarefaction and a right shock. Test 2 has a solution consisting of two rarefaction waves, while Test 3 consists of two shocks. We consider two different values of the speed of sound: $c_s=10^{-3}$ and $10^{-4}$. We set the initial discontinuity at $x_0=1$ and sample the solution in the interval $0<x<2$ at time $\tau=600$. We set the accuracy of the Newton-Raphson method to $\epsilon=10^{-6}$. 

The exact solutions to Test 1, 2 and 3 are shown in Figs.~\ref{fig2},~\ref{fig3} and~\ref{fig4} respectively. In the case of Test 1 we can clearly see the opening of the rarefaction fan moving from right to left and the shock wave propagating to the right. For Test 2 we notice the opening of the two rarefaction fans propagating to the left and to the right respectively, while in the case of Test 3 we can see the density and velocity profile of the colliding shock waves. It is worth noting that these solutions only depend on the value of the sound speed, and are independent of the equation of state.

\subsection{Approximate Riemann Solvers}\label{appriemann}
While the Riemann problem can be solved using an exact solver as the one described in Section~\ref{exact_riemann}, for any practical application such as cosmological simulations this is computationally too expensive since it requires solving iteratively algebraic equations at each cell of the spatial domain of integration. For this reason, in this section, we present some approximate solvers. Let us notice here that in cosmological simulations one can combine different Riemann solvers in an adaptive fashion, where the more accurate and more expensive solvers are only used in case of large gradients, while the cheaper ones are used for smooth flow configurations, which represent the vast majority of the field.
\subsubsection{Primitive Variables Riemann Solver}\label{PVRS}
Let us consider the characteristic equations of the advection equation Eq.~(\ref{advprimiv}) in primitive variables:
\begin{equation}
  \tilde{\textbf{L}}_\pm \cdot d\textbf{W}=0,
\end{equation}
where $\tilde{\textbf{L}}_\pm$ are the left eigenvectors Eq.~(\ref{lefteigenvec}) and $d\textbf{W}=(d\Pi,dv)$. These read: 
\begin{eqnarray}
\frac{d\Pi}{\Pi}&+&\frac{1+\frac{c_s^2}{c^2}}{c_s}\frac{dv}{\sqrt{1-\frac{v^2}{c^2}}+\frac{v c_s}{c^2}}=0 \quad \,\textrm{along} \,\,\, dx/d\tau=\lambda_+\label{charp}\\
\frac{d\Pi}{\Pi}&-&\frac{1+\frac{c_s^2}{c^2}}{c_s}\frac{dv}{\sqrt{1-\frac{v^2}{c^2}}-\frac{v c_s}{c^2}}=0 \quad \,\textrm{along} \,\,\, dx/d\tau=\lambda_-.\label{charm}
\end{eqnarray}
Following the derivation presented in \cite{Toro2009}, we can connect the star state to the left and right states by integrating these equations along the direction of the characteristics. We can simplify the solution by defining the quantities:
\begin{eqnarray}
C^+&=&\frac{1+\frac{c_s^2}{c^2}}{c_s}\frac{\Pi}{\sqrt{1-\frac{v^2}{c^2}}+\frac{v c_s}{c^2}},\\
C^-&=&\frac{1+\frac{c_s^2}{c^2}}{c_s}\frac{\Pi}{\sqrt{1-\frac{v^2}{c^2}}-\frac{v c_s}{c^2}}.
\end{eqnarray}
and approximating them as constant in time. This means that we can compute their value at the foot of the characteristic by only using the known left and right state variables. We can then connect the star state to the left (right) state by integrating Eq.~(\ref{charp}) (Eq.(\ref{charm})) along the characteristic of speed $\lambda_+$ ($\lambda_-$). This gives:
\begin{eqnarray}
\Pi_L+C_{+,L}v_L=\Pi_*+C_{+,L}v_*, \label{foot1}\\
\Pi_*-C_{-,R}v_*=\Pi_R-C_{-,R}v_R. \label{foot2}
\end{eqnarray}
Solving the linear system of Eqs.~(\ref{foot1}) and~(\ref{foot2}) we finally obtain the approximate state variables in the star region:
\begin{eqnarray}
\Pi_*&=\frac{C_{+,L}\Pi_R+C_{-,R}\Pi_L+C_{-,R}C_{+,L}(v_L-v_R)}{C_{-,R}+C_{+,L}} \\
v_*&=\frac{\Pi_L-\Pi_R+C_{+,L}v_L+C_{-,R}v_R}{C_{-,R}+C_{+,L}}
\end{eqnarray}
We refer to these solutions as the Linear Primitive Variable Riemann Solver (LPVRS).

An alternative approximation can be found by defining:
\begin{eqnarray}
C^+&=&\frac{1+\frac{c_s^2}{c^2}}{c_s}\frac{1}{\sqrt{1-\frac{v^2}{c^2}}+\frac{v c_s}{c^2}},\\
C^-&=&\frac{1+\frac{c_s^2}{c^2}}{c_s}\frac{1}{\sqrt{1-\frac{v^2}{c^2}}-\frac{v c_s}{c^2}}.
\end{eqnarray}
Integrating Eq.~(\ref{charp}) (Eq.~(\ref{charm})) along the characteristic of speed $\lambda_+$ ($\lambda_-$) in this case gives:
\begin{eqnarray}
\ln{\Pi_{\rm L}}-\ln{\Pi^*}=-C^+_{\rm L} v_{\rm L} + C^+_{\rm L} v^*,\label{foot1b} \\
\ln{\Pi^*}-\ln{\Pi_{\rm R}}=C^-_{\rm R} v^* - C^-_{\rm R} v_{\rm R},\label{foot2b}
\end{eqnarray}
and the approximate state variables in the star region are:
\begin{eqnarray}
v^*&=&\frac{C^+_{\rm L}v_{\rm L}+C^-_{\rm R}v_{\rm R}+\ln{\Pi_{\rm L}}-\ln{\Pi_{\rm R}}}{C^+_{\rm L}+C^+_{\rm R}},\\
\Pi^*&=&\exp{\left[\frac{C^+_{\rm L}\ln{\Pi_{\rm R}}+C^-_{\rm R}\ln{\Pi_{\rm L}}+C^+_{\rm L}C^-_{\rm R}(v_{\rm L}-v_{\rm R})}{C^+_{\rm L}+C^+_{\rm R}}\right]}.
\end{eqnarray}
We refer to this second approximation as the Primitive Variable Riemann Solver (PVRS). Notice that, unlike the LPVRS case, we no longer approximate $\Pi$ as being constant along the characteristic.

To find a simple solution for the state variables inside the rarefaction fan we make the approximation that $\sqrt{1-\frac{v^2}{c^2}}\approx1$ and expand the functions $g_{\rm Lr}$ and $g_{\rm Rr}$ given by Equations \ref{glr_def} and \ref{grr_def} to first order. This gives:
\begin{eqnarray}
    v_{\rm L}^{\rm fan}&=&\frac{x-x_0}{\tau}+c_s \\
    \Pi_{\rm L}^{\rm fan}&=&\exp{\left(\log{\Pi_L}+\frac{v_{\rm L}}{c_s}+\frac{v_{\rm L}c_s}{c^2}-\frac{v_{\rm L}^{\rm fan}}{c_s}-\frac{v_{\rm L}^{\rm fan}c_s}{c^2}\right)}\\
    v_{\rm R}^{\rm fan}&=&\frac{x-x_0}{\tau}-c_s \\
    \Pi_{\rm R}^{\rm fan}&=&\exp{\left(\log{\Pi_R}-\frac{v_{\rm R}}{c_s}-\frac{v_{\rm R}c_s}{c^2}+\frac{v_{\rm R}^{\rm fan}}{c_s}+\frac{v_{\rm R}^{\rm fan}c_s}{c^2}\right)}
\end{eqnarray}

\begin{figure}[ht]
\centering
\begin{tabular}{cc}
\includegraphics[width=5.5cm]{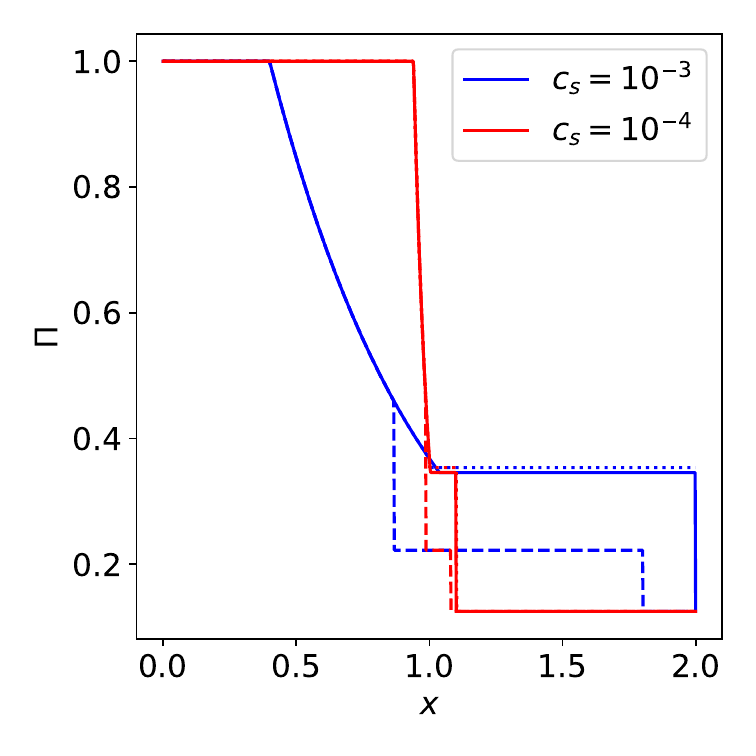} &
\includegraphics[width=5.5cm]{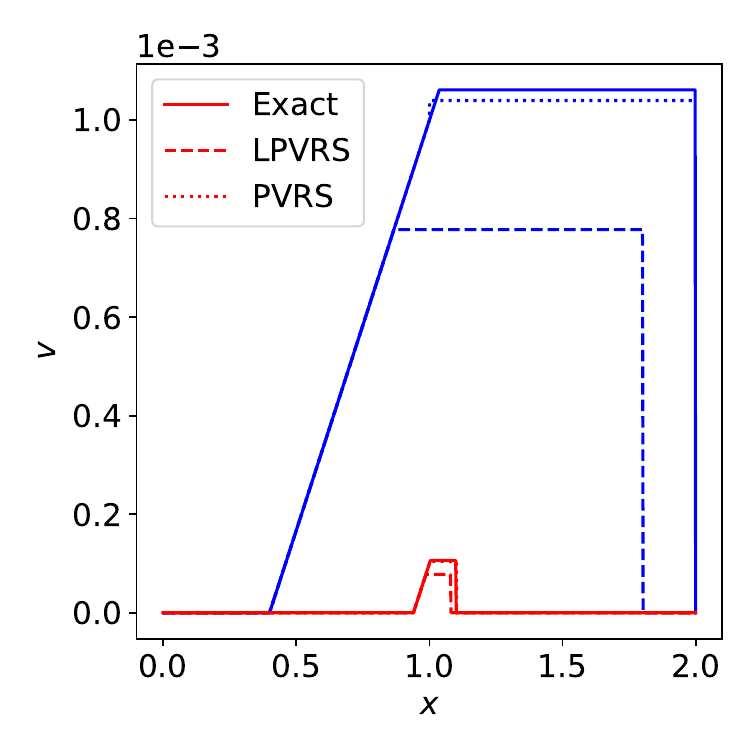} \\
\end{tabular}
\caption{Solution of Test 1 for $\Pi$ (left panel) and $v$ (right panel) at time $\tau=600$ for $c_s=10^{-3}$ (blue) and $10^{-4}$ (red) respectively for the exact solver (continuous line), LPVRS (dashed line) and PVRS (dotted line).\linebreak}\label{fig_test1}
\centering
\begin{tabular}{cc}
\includegraphics[width=5.5cm]{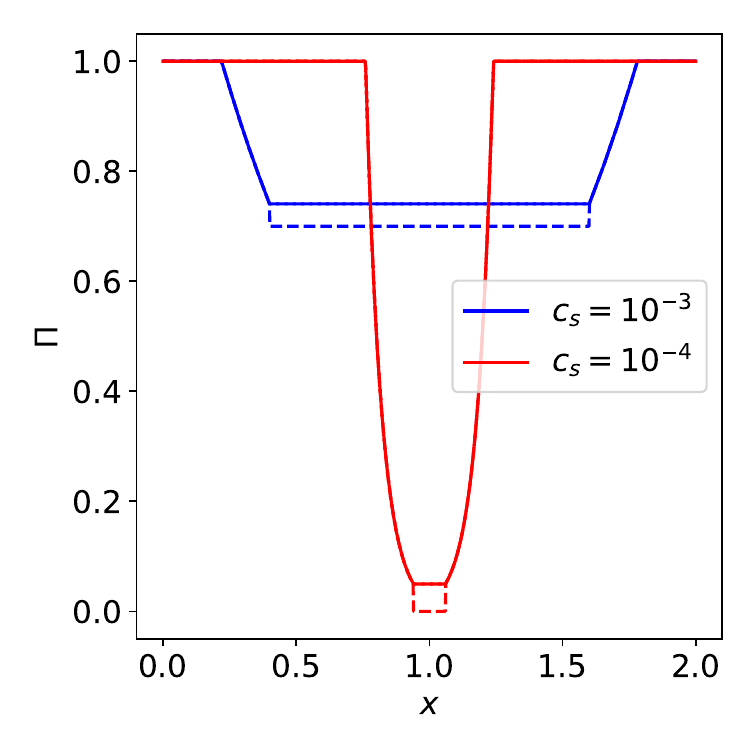}&
\includegraphics[width=5.5cm]{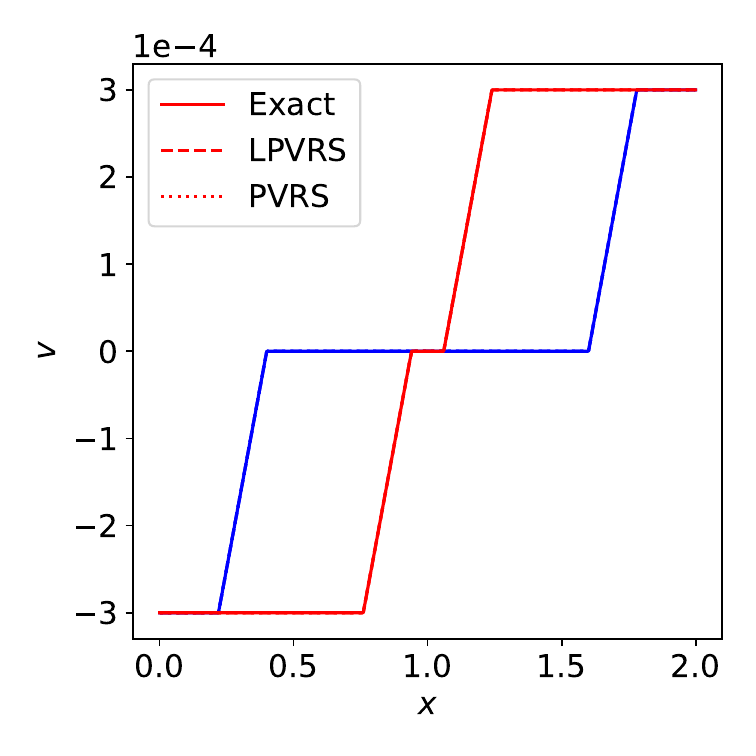} \\
\end{tabular}
\caption{As in Fig.~\ref{fig_test1} for Test 2.\linebreak}\label{fig_test2}
\centering
\begin{tabular}{cc}
\includegraphics[width=5.5cm]{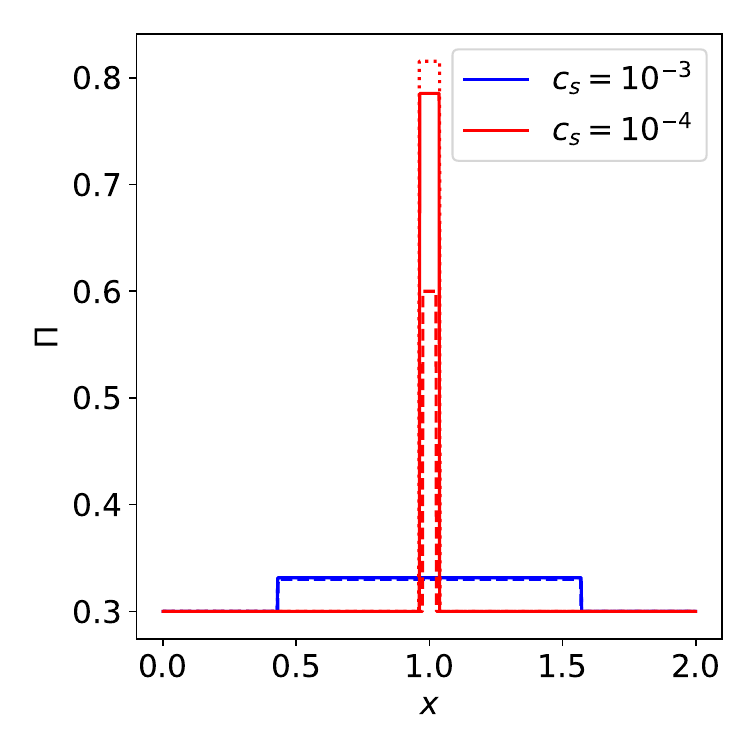}&
\includegraphics[width=5.5cm]{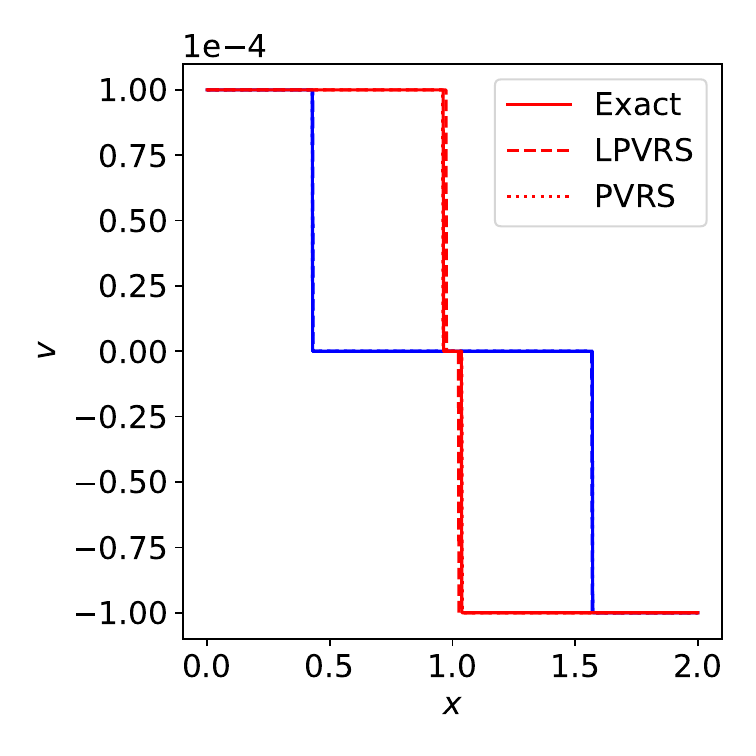} \\
\end{tabular}
\caption{As in Fig.~\ref{fig_test1} for Test 3.}\label{fig_test3}
\end{figure}

We compare the results of these approximations to the exact solution for the tests in Table~\ref{tab1} in Figures \ref{fig_test1}-\ref{fig_test3}. We can see that the PRVS  performs very well in the case of rarefactions, where the solution is indistinguishable from the exact one, and only slightly misestimates $\Pi^*$ in the presence of shock waves. On the other hand, the LPVRS gives a larger error on the value of $\Pi^*$ in all the tests and on the value of $v^*$ in Test 1, especially for the larger value of $c_s$. While the PVRS approximation gives better results with respect to the LPVRS in all the tests, it is however susceptible to numerical instabilities, given its non-linear nature.

\subsubsection{Harten, Lax and van Leer Riemann Solver}\label{HLL}
The Harten, Lax and van Leer (HLL) Riemann solver approximates the solution as a system of two waves that separate three constant states. The intermediate state is found by averaging the exact solution between the fastest waves that travel to the left and right of the discontinuity \citep[for a more detailed explanation see][]{Toro2009}.

Given an estimate of these velocities $s_{\rm L}$ and $s_{\rm R}$, the HLL solver approximates the flux directly as:
\begin{equation}
\textbf{F}=
    \begin{cases}
    \textbf{F}_{\rm L} \quad \text{if }x-x_0\le s_{\rm L}\tau\\
    \textbf{F}_{\rm HLL} \quad \text{if }s_{\rm L}\tau\le x-x_0\le s_{\rm R}\tau\\
    \textbf{F}_{\rm R} \quad \text{if }x-x_0\ge s_{\rm R}\tau
    \end{cases}
\end{equation}
where
\begin{equation}
\textbf{F}_{\rm HLL}=\frac{s_{\rm R} \textbf{F}_{\rm L} - s_{\rm L} \textbf{F}_{\rm R} + s_{\rm L} s_{\rm R} (\textbf{U}_{\rm R} - \textbf{U}_{\rm L})}{s_{\rm R} - s_{\rm L}}.
\end{equation}
The main advantage of this solver is its efficiency since it gives an estimate of the flux that involves only a few basic operations.

\begin{figure}[ht]
\centering
\begin{tabular}{cc}
\includegraphics[width=5.5cm]{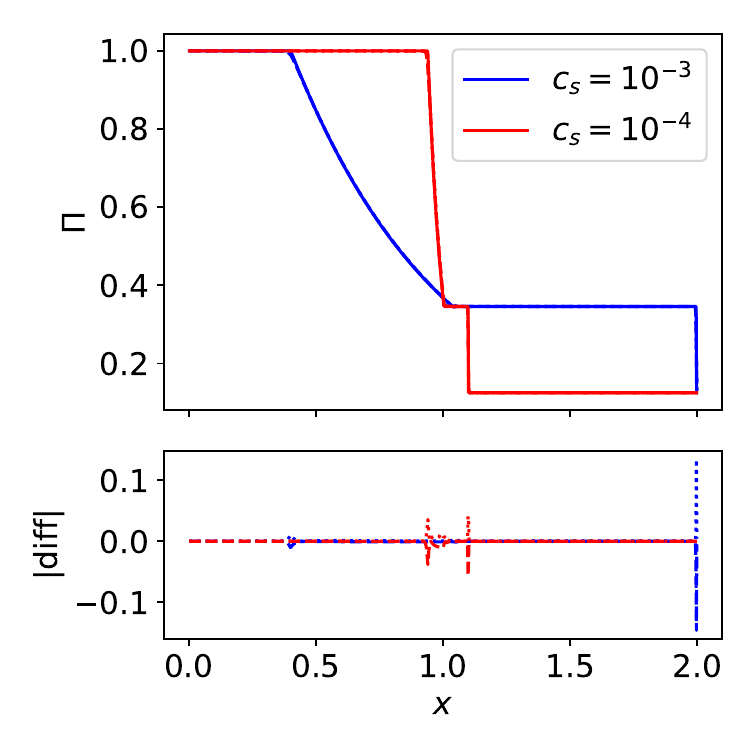} &
\includegraphics[width=5.5cm]{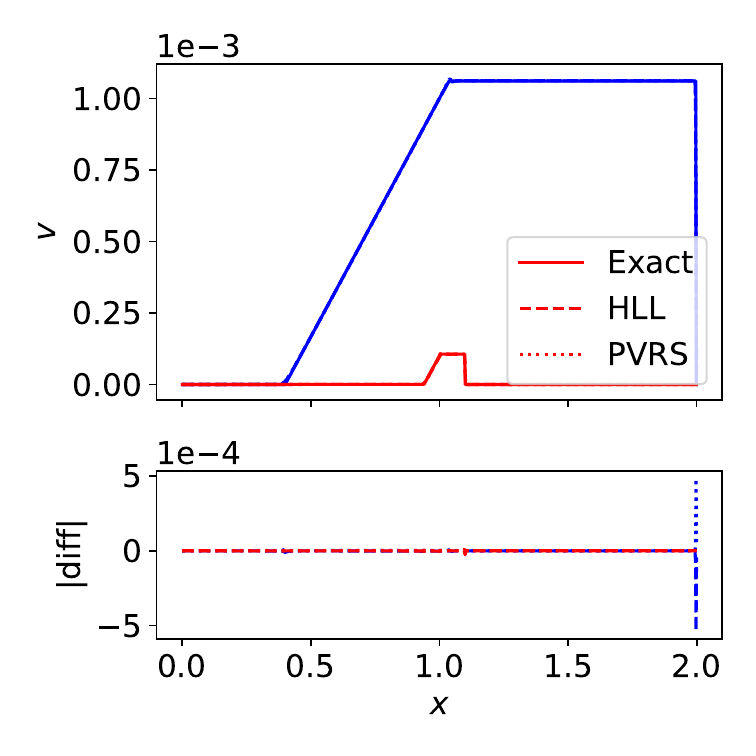} \\
\end{tabular}
\caption{Solution of Test 1 for $\Pi$ (left panel) and $v$ (right panel) at time $\tau=600$ for $c_s=10^{-3}$ (blue) and $10^{-4}$ (red) for the exact Riemann solver (continuous line) and the MUSCL-Hancock method with PVRS (dotted line) and HLL Riemann solver (dashed line) respectively. Bottom panels: the absolute difference between the numerical and the exact solution with opposite signs for the two Riemann solvers to better see the distinction.\linebreak}\label{fig_test1_hll}
\end{figure}

\begin{figure}
\centering
\begin{tabular}{cc}
\includegraphics[width=5.5cm]{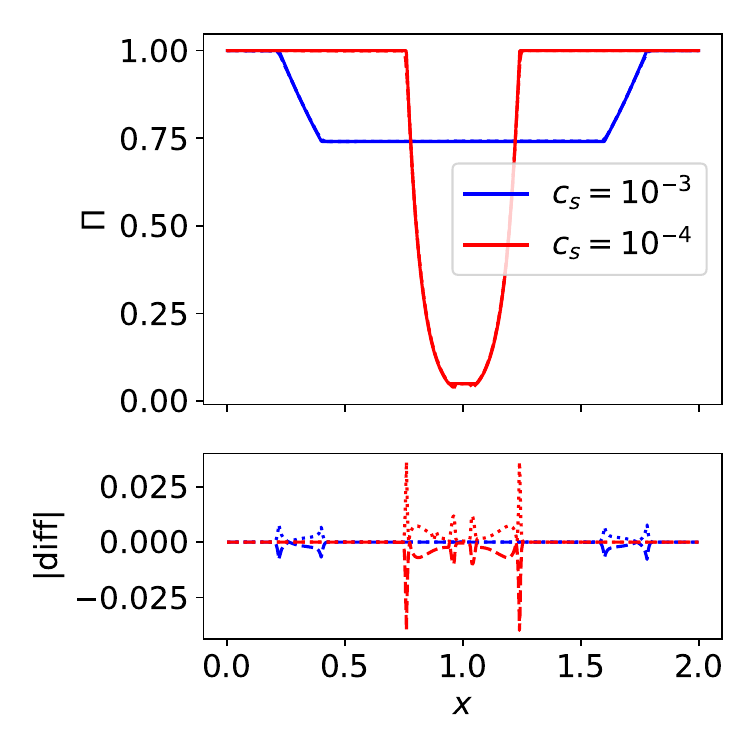}&
\includegraphics[width=5.5cm]{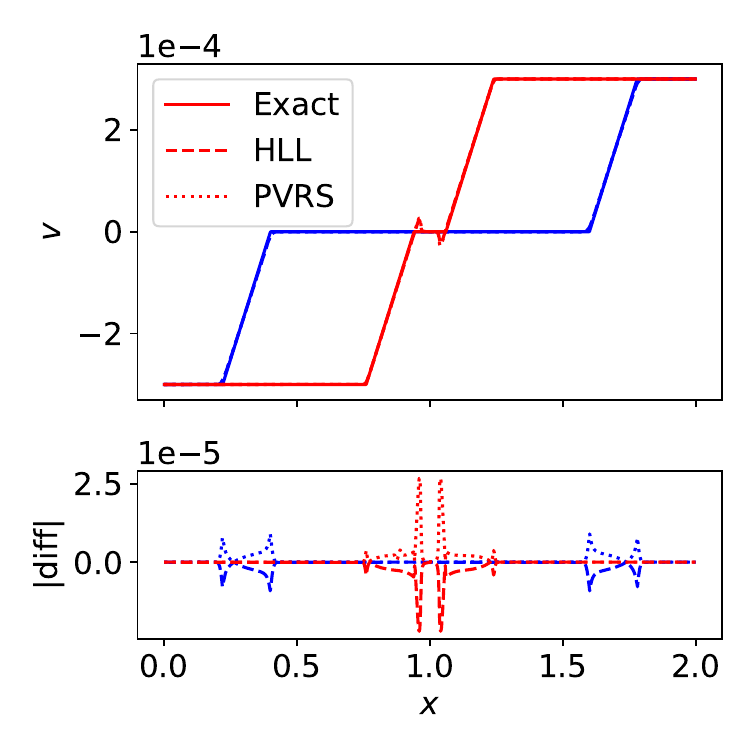} \\
\end{tabular}
\caption{As in Fig.~\ref{fig_test1_hll} for Test 2.\linebreak}\label{fig_test2_hll}
\centering
\begin{tabular}{cc}
\includegraphics[width=5.5cm]{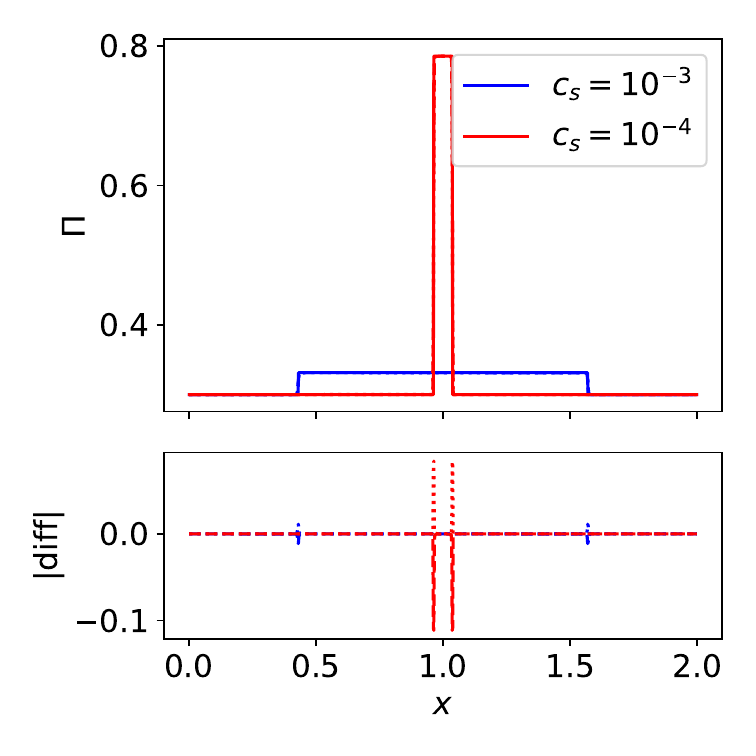}&
\includegraphics[width=5.5cm]{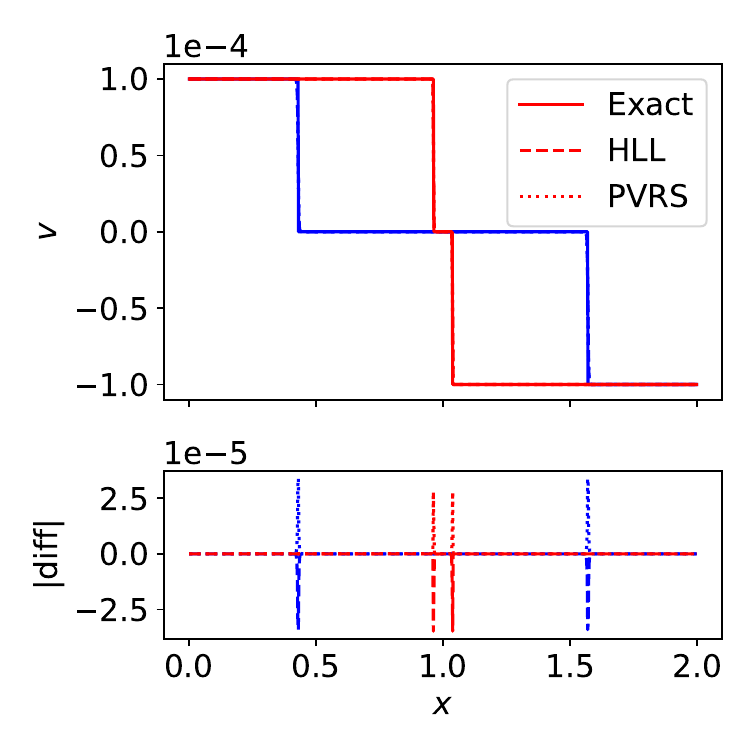} \\
\end{tabular}
\caption{As in Fig.~\ref{fig_test1_hll} for Test 3.}\label{fig_test3_hll}
\end{figure}

To test this solver we employ the simple estimates:
\begin{eqnarray}
    s_{\rm L}={\rm min}\left\{ v_{\rm L}-c_s\sqrt{1-\frac{v_{\rm L}^2}{c^2}}, v_{\rm R}-c_s\sqrt{1-\frac{v_{\rm R}^2}{c^2}}\right\}, \\
    s_{\rm R}={\rm max}\left\{ v_{\rm L}+c_s\sqrt{1-\frac{v_{\rm L}^2}{c^2}}, v_{\rm R}+c_s\sqrt{1-\frac{v_{\rm R}^2}{c^2}}\right\}.
\end{eqnarray}
We verified that using more complex estimates for the speed does not significantly improve the accuracy of the solution, so that it is not worth spoiling the efficiency of this approximation. Since this solver gives an approximation for the flux, its performance is best shown using a Godunov solver. We compare the results of the tests in Table~\ref{tab1} using the MUSCL-Hancock method described in Section~\ref{muscl} with PVRS and HLL Riemann solvers in Figures~\ref{fig_test1_hll}-\ref{fig_test3_hll}. We can see that the HLL solver performs almost identically to PVRS in these tests.


\section{Finite Volume Conservative Methods}\label{upwind}
Let us consider the problem of solving Eq.~(\ref{vec_eq}) numerically, provided with a set of initial and boundary conditions. The first step consists in discretizing the time interval into a finite number of time-steps $\tau^n$ of size $\Delta\tau$ and the spatial domain in a finite number of cells $I_i=[x_{i-\frac{1}{2}},x_{i+\frac{1}{2}}]$ of size $\Delta x=L/M$, where $i=1,...,M$ and $L$ is the size of the spatial domain of integration. This choice of spatial discretization is what characterizes finite volume methods, where the state $\textbf{U}_i$ is regarded as a volume average of the variable $\textbf{U}$ over the cell $I_i$. In the following we will denote the numerical approximation of $\textbf{U}$ in the cell $I_i$ at time $\tau^n$ as $\textbf{U}^n_i$.

A numerical solution of Eq.~(\ref{vec_eq}) can be obtained by using the operator splitting approach, where the system is split into a purely advection problem with $\textbf{U}_\tau+\textbf{F}(\textbf{U})_x=0$ and an ordinary differential equation (ODE) $\textbf{U}_\tau=\textbf{S}(\textbf{U})$. 
In such a case, one solves the PDE associated to the advection equations and then uses the solution to solve the ODE with a method that matches the order of accuracy of the solution of the advection problem such as to retain the same level of accuracy globally. In the following we focus on the solution of the advection equations, while the solution to the ODE will be implemented into the full cosmological simulation code that we will present in future work.

Following standard textbook presentations \citep[see e.g.][]{Toro2009}, the numerical solution to the advection problem can be obtained using a conservative scheme, such as that introduced by Godunov \citep{Godunov1959}:
\begin{equation}
\textbf{U}_i^{n+1}=\textbf{U}_i^{n}+\frac{\Delta \tau}{\Delta x}\left[\textbf{F}_{i-\frac{1}{2}}-\textbf{F}_{i+\frac{1}{2}}\right],\label{conservative}
\end{equation}
where the $\textbf{F}_{i\pm\frac{1}{2}}$ are inter-cell fluxes at the boundaries of the $i$-th cell given
by
\begin{equation}
\textbf{F}_{i\pm\frac{1}{2}}=\textbf{F}_{i\pm\frac{1}{2}}[\textbf{U}_{i\pm 1/2}(0)],
\end{equation}
where 
$\textbf{U}_{i+1/2}(0)$ is the solution of the Riemann problem $RP(\textbf{U}_i^n,\textbf{U}_{i+1}^n)$ at $x/t=0$ and $\textbf{U}_{i-1/2}(0)$ is the solution of the Riemann problem $RP(\textbf{U}_{i-1}^n,\textbf{U}_{i}^n)$ at $x/t=0$. The size of the time-step $\Delta\tau$ satisfies the condition $\Delta\tau\le\Delta x/s^n_{\rm max}$, where $s^n_{\rm max}$ is the maximum wave velocity at time $\tau^n$ in the spatial domain of integration. This condition ensures that no wave in the solution of the Riemann problem travels more than the size of the cell $\Delta x$ in the time interval $\Delta \tau$. A time step satisfying the above condition can be set by introducing the Courant-Friedrichs-Lewy (CFL) coefficient $C_{\rm cfl}$, such that 
\begin{equation}
\Delta\tau=C_{\rm cfl}\frac{\Delta x}{s^n_{\rm max}},
\end{equation}
with $0<C_{\rm cfl}<1$.

Notice that using piecewise constant data as initial condition for the Riemann problem, this method gives solutions that are only first order accurate in space and time. However, higher-order accuracy in space and/or time can be achieved by reconstructing the state variables using a truncated Taylor expansion and then use the reconstructed state as input for the Riemann solver. Here, we will derive working examples of some of these higher-order schemes such as the Monotonic Upstream-Centered Scheme for Conservation Laws (MUSCL-Hancock), the Piecewise Linear Method (PLM) and the Piecewise Parabolic Methods (PPM).

We test these schemes against the exact solutions of the Sod, the two rarefaction, and the two shock waves tests, which we have presented in Section~\ref{riemann_solvers}. Moreover, as we expect some of these schemes to improve the calculation of the flow in smooth regions rather than at discontinuities, we perform a pure advection test, which we discuss in Section~\ref{advection_test}.

Hereafter, we use the Primitive Variables Riemann Solver (PVRS) described in Section~\ref{PVRS}. We set $C_{\rm cfl}=0.9$ and estimate the maximum wave velocity as
\begin{equation}
s^n_{\rm max}=\max_{i}\{|v_i^n|+c_s\},
\end{equation}
which provides a very good approximation in the case of rarefaction waves, though it may underestimate the speed of propagation in the case of shocks.

We assume periodic boundary conditions for the advection test:
\begin{equation}
\Pi_{1}^n=\Pi_{M}^n\,,  \quad v_{1}^n=v_{M}^n,
\end{equation}
while for the other test we assume transmissive boundary conditions:
\begin{equation}
\Pi_{M+1}^n=\Pi_{M}^n\,,  \quad v_{M+1}^n=v_{M}^n
\end{equation}
so that the boundaries do not affect the propagation of waves.

\subsection{MUSCL-Hancock Method}\label{muscl}
As already mentioned, Godunov's original scheme, which is based on a piecewise constant distribution of data, is  first-order accurate in both space and time. In a series of seminal papers \cite{VanLeer1973,VanLeer1974,VanLeer1977a,VanLeer1977a,VanLeer1977b,VanLeer1979} Van Leer presented a modification of Godunov's method to achieve higher-order accuracy. The idea is to use the initial piecewise constant data to reconstruct the distribution of state variables inside cells and extrapolate their values at the cell interfaces. Then, using the conservation equations these boundary values are evolved by half time-step to achieve second-order accuracy in time. It is these reconstructed values that provide the initial data for the Riemann problem at the cell interface. While first-order schemes are guaranteed to preserve the monotonicity of the solution by Godunov's theorem, this is not the case for higher-order methods. In fact, the data reconstruction inside cells may generate local extrema that lead to spurious numerical oscillations in the solution. We thus need to impose monotonicity constraints on the reconstructed states, as we will see below.

The MUSCL-Hancock method introduced by Van Leer\footnote{In \cite{VanLeer1984} the idea behind this scheme is attributed to S. Hancock, see also \cite{VanLeer2006}, from which the acronym of MUSCL-Hancock method.} \cite{VanLeer1984} is based on a linear reconstruction of the data distribution inside cells. The algorithm consists of the following step:
\begin{itemize}
\item Starting with the initial piecewise constant data $\textbf{U}_i^{n}$, the left and right cell boundary value are linearly extrapolated as
\begin{equation}
\textbf{U}_{i,L}^n=\textbf{U}_i^{n}-\frac{1}{2}\partial\textbf{U}_i\,,  \quad \textbf{U}_{i,R}^n=\textbf{U}_i^{n}+\frac{1}{2}\partial\textbf{U}_i\,,
\end{equation}
where $\partial\textbf{U}_i$ denotes the slope of the linear interpolation in the $i$-th cell, whose determination will be described below.
\item Evolve $\textbf{U}_{i,L}^n$ and $\textbf{U}_{i,R}^n$ by a time $\Delta \tau/2$ using the conservative update 
\begin{eqnarray}
{\rm\bar{\bf U}}_{i,L}^{n+1/2}&=&\textbf{U}_{i,L}^n+\frac{1}{2}\frac{\Delta \tau}{\Delta x}\left[\textbf{F}(\textbf{U}_{i,L}^n)-\textbf{F}(\textbf{U}_{i,R}^n)\right],\\
{\rm\bar{\bf U}}_{i,R}^{n+1/2}&=&\textbf{U}_{i,R}^n+\frac{1}{2}\frac{\Delta \tau}{\Delta x}\left[\textbf{F}(\textbf{U}_{i,L}^n)-\textbf{F}(\textbf{U}_{i,R}^n)\right].
\end{eqnarray}
\item Solve the Riemann Problem at inter-cell boundary with piecewise constant data ${\rm\bar{\bf U}}_{i,L}^{n+1/2}$ and ${\rm\bar{\bf U}}_{i,R}^{n+1/2}$:
\begin{equation}
\textbf{U}_\tau+\textbf{F}(\textbf{U})_x=0,
\end{equation}
with initial conditions
\begin{equation}
\textbf{U}(x,0)=
    \begin{cases}
    {\rm\bar{\bf U}}_{i,L}^{n+1/2} & \quad \text{if $x<0$}\\
    {\rm\bar{\bf U}}_{i+1,R}^{n+1/2} & \quad \text{if $x>0$}
    \end{cases} 
\end{equation}
\end{itemize}

A key point concerns the evaluation of the slopes $\partial\textbf{U}_i$, which must satisfy monotonicity constraints. We implement the MUSCL-Hancock scheme with the MINMAX slope limiter which reads as:
\begin{equation}
\partial\textbf{U}_i=\frac{1}{2}\left[\textrm{sign}(\textbf{U}_i^{n}-\textbf{U}_{i-1}^{n})+\textrm{sign}(\textbf{U}_{i+1}^{n}-\textbf{U}_{i}^{n})\right]\times\textrm{min}\left(|\textbf{U}_i^{n}-\textbf{U}_{i-1}^{n}|,|\textbf{U}_{i+1}^{n}-\textbf{U}_{i}^{n}|\right).
\end{equation}

\begin{figure}[th]
\centering
\begin{tabular}{cc}
\includegraphics[width=5.5cm]{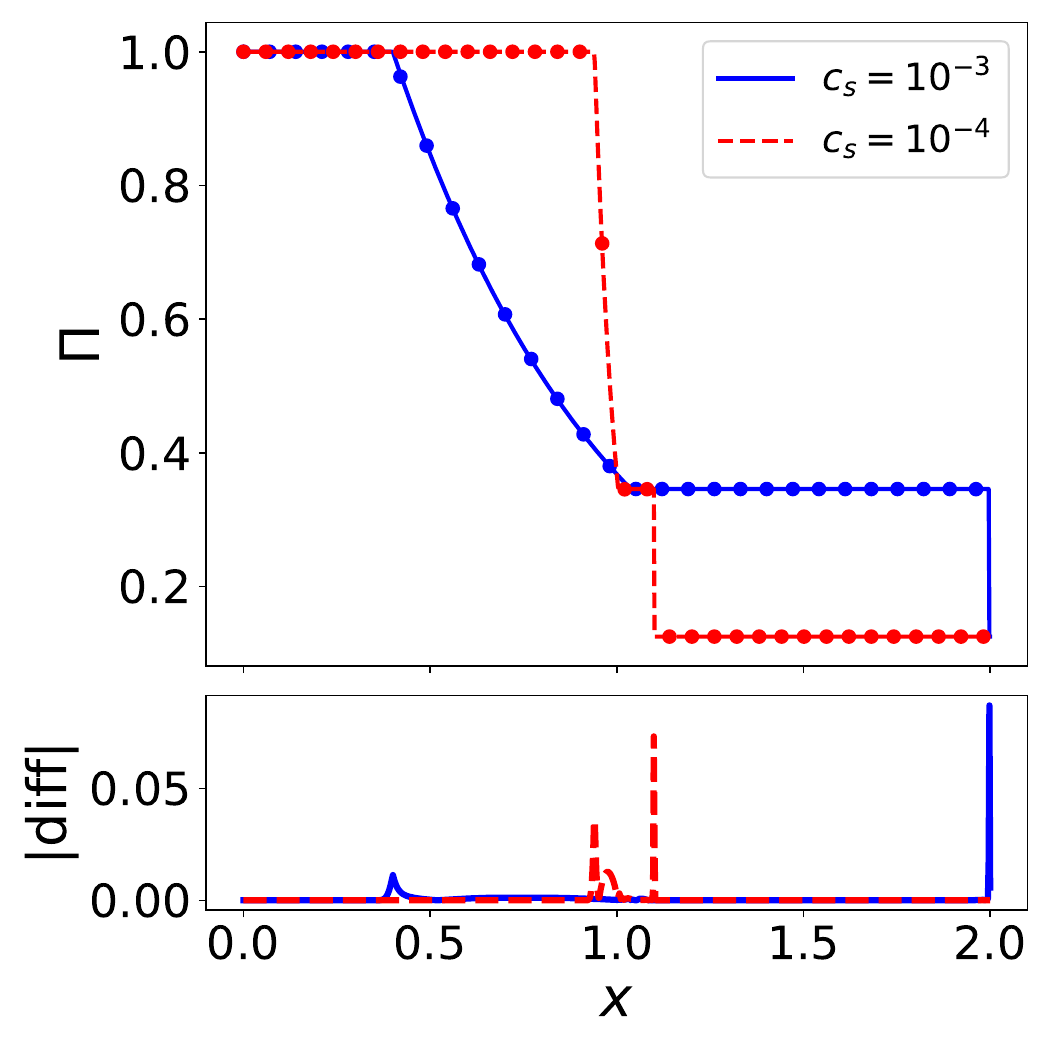} &
\includegraphics[width=5.5cm]{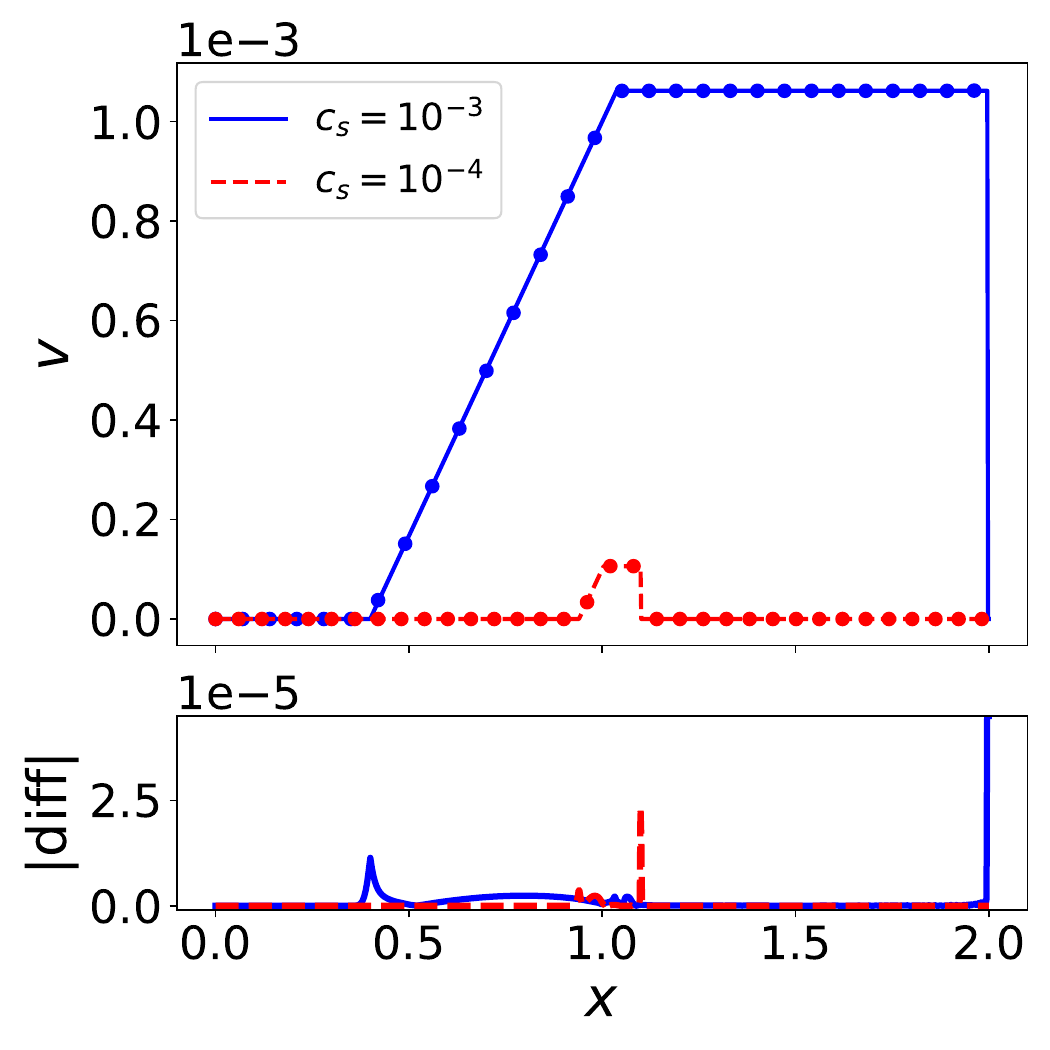} \\
\end{tabular}
\caption{MUSCL-Hancock (circles) vs. Exact solution (lines) of Test 1 for $\Pi$ (left panel) and $v$ (right panel) at time $\tau=600$ units for $c_s=10^{-3}$ (blue solid line) and $10^{-4}$ (red dashed line) respectively. In the bottom panel the absolute difference between the numerical and exact solution.\linebreak}\label{fig5}
\end{figure}
\begin{figure}[h]
\centering
\begin{tabular}{cc}
\includegraphics[width=5.5cm]{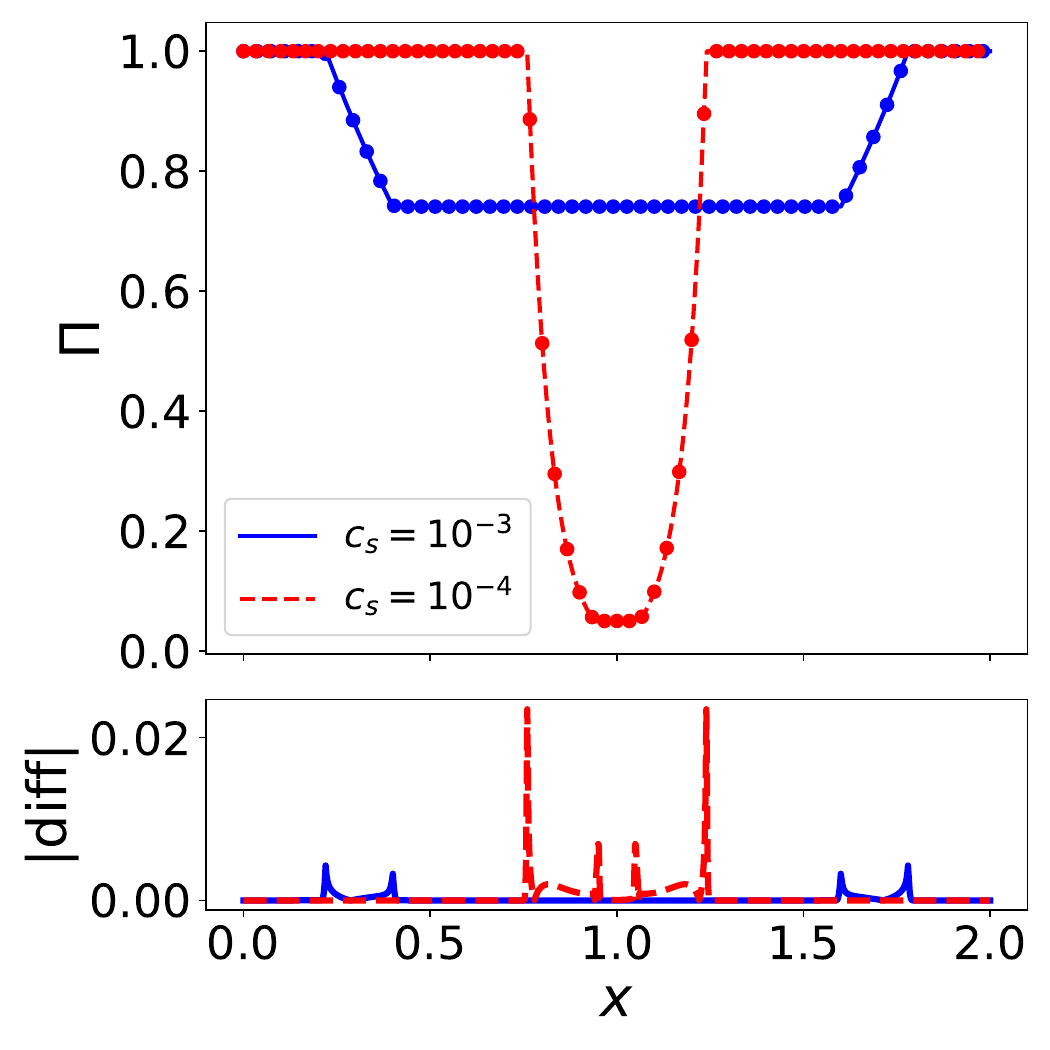}&
\includegraphics[width=5.5cm]{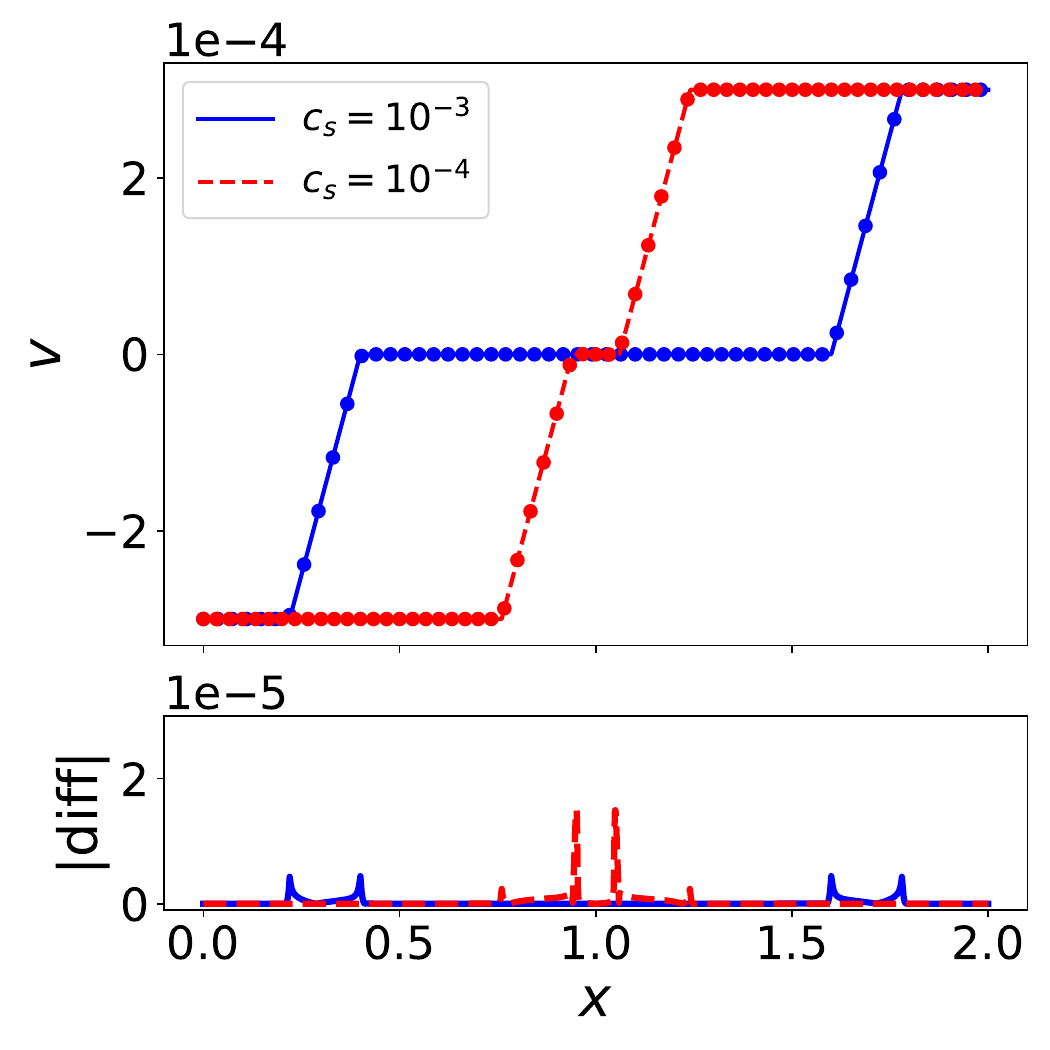} \\
\end{tabular}
\caption{As in Fig.~\ref{fig5} for Test 2.\linebreak}\label{fig6}
\end{figure}
\begin{figure}[h]
\centering
\begin{tabular}{cc}
\includegraphics[width=5.5cm]{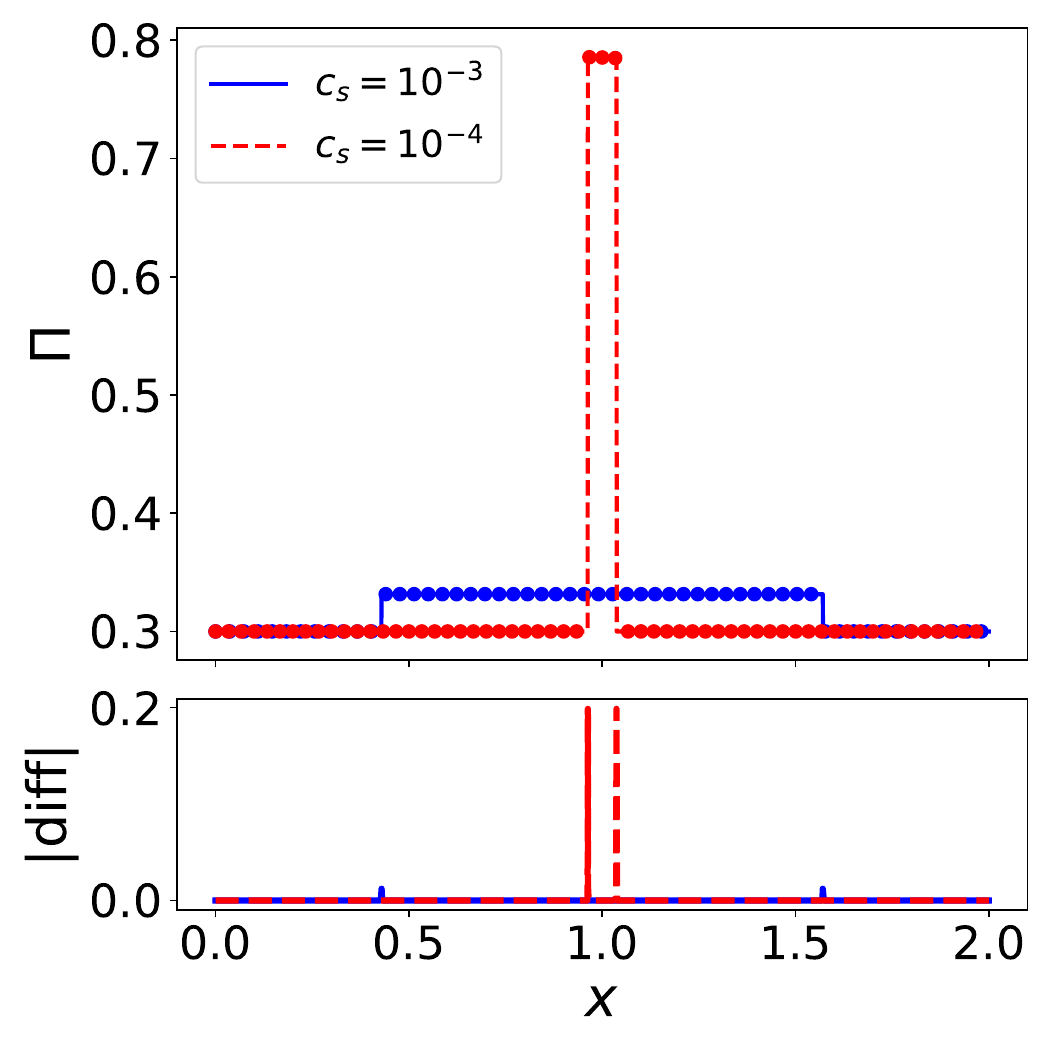}&
\includegraphics[width=5.5cm]{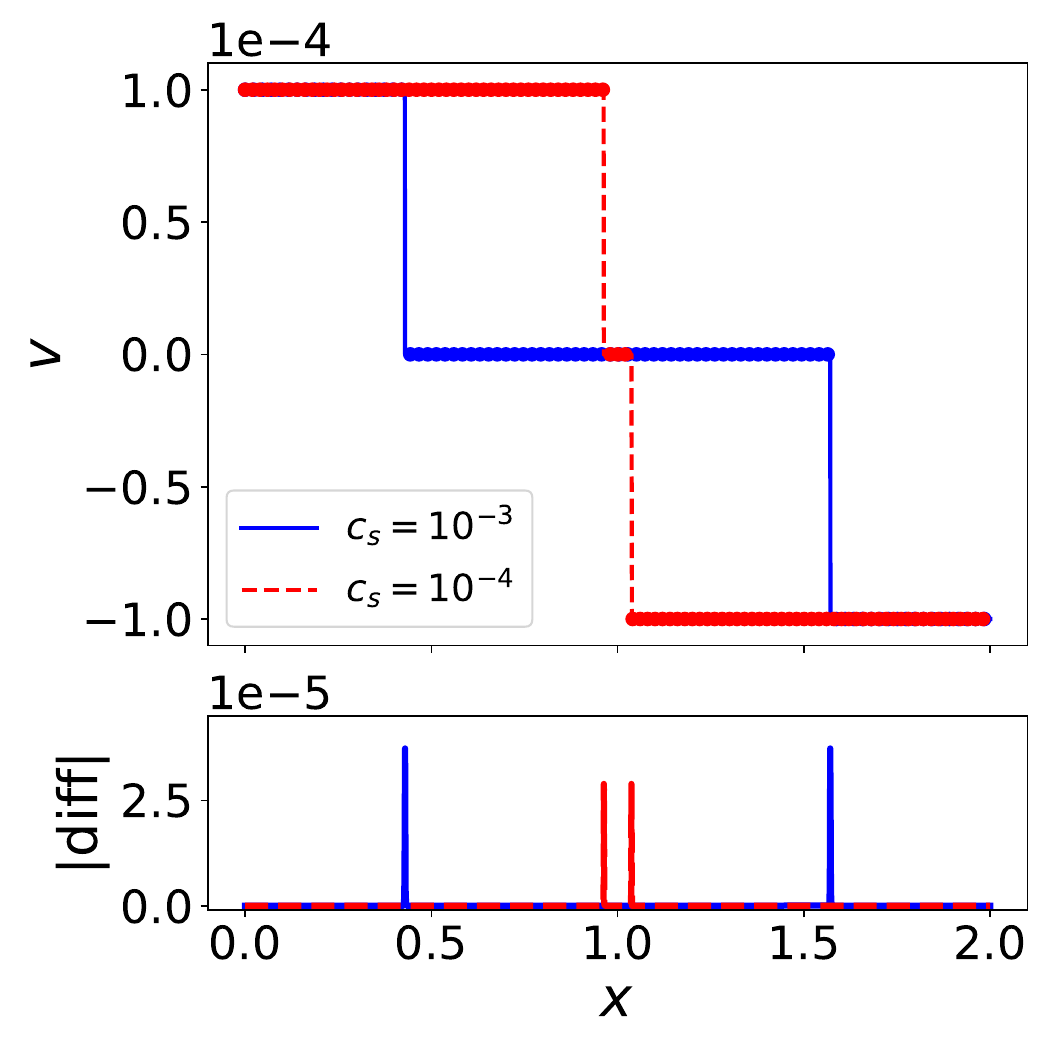} \\
\end{tabular}
\caption{As in Fig.~\ref{fig5} for Test 3.}\label{fig7}
\end{figure}

In Figs.~\ref{fig5},~\ref{fig6} and~\ref{fig7} we plot the numerical solutions (dotted lines) to test cases given in Table~\ref{tab1} obtained with the MUSCL-Hancock method against the exact solutions at $\tau=600$ units for $c_s=10^{-3}$ (solid blue lines) and $c_s=10^{-4}$ (dash red lines) respectively. In the bottom panels we plot the absolute value of the difference between the numerical and exact solutions.

Overall the scheme performs rather well. A zoom on the value of $\Pi$ near the shocks in Figs.~\ref{fig5} and ~\ref{fig7} reveals that shock waves are spread over $\sim 3$ cells, rather than having the zero-width of the exact solution. On the other hand, the speed is correctly estimated as can be see on the right panels, thus indicating that the average shock position is well determined by the numerical scheme. Another characteristic feature of the solutions of Test 1 and 3 near the shocks is the absence of spurious oscillations. In the case of rarefaction waves, larger errors occur near the head and the tail as can be seen from Figs.~\ref{fig5} and~\ref{fig6}, discontinuities are spread over $\sim 10$ cells. The numerical solution converges to the exact by increasing the spatial resolution of the grid.

\subsection{Piecewise Linear Method (PLM)}\label{PLM}
This scheme, introduced by Colella \cite{Colella1985}, builds upon the MUSCL method and uses a linear reconstruction of the data to achieve second-order accuracy in both space and time. We will follow the implementation presented in the lecture notes by M. Zingale \cite{Zingale} to which we refer the reader for a clear summary of higher-order reconstruction methods of non-linear advection equations. 

The basic idea is to obtain left and right inter-cell states as a Taylor expansion of the primitive variables to first-order in $\Delta x/2$ to displace from the cell-centered value to the interface, and to first-order in $\Delta \tau/2$ to evolve the states to mid-point in time. A crucial aspect of this scheme is the so-called \textit{characteristic tracing} step, where only the waves that travel towards the interface are selected during the reconstruction process. This ensures that the correct flow of information is preserved. 

These states are constructed as:
 \begin{equation}
\bar{\Pi}_{i,L}^{n+1/2}=\Pi_i^{n}+\frac{1}{2}\left[\tilde{\Pi}_{i}^{-}+\tilde{\Pi}_{i}^{+}\right]\,,  \quad \bar{\Pi}_{i+1,R}^{n+1/2}=\Pi_{i+1}^{n}-\frac{1}{2}\left[\tilde{\Pi}_{i+1}^{-}+\tilde{\Pi}_{i+1}^{+}\right] \,,
\end{equation}
with
\begin{equation}
\tilde{\Pi}_{i}^{\pm}=
    \begin{cases}
    \left(1-\frac{\Delta \tau}{\Delta x}\lambda_{\pm}(v_i)\right)\left(\tilde{L}_{1,i}^{\pm}\,\tilde{R}_{1,i}^{\pm}\partial\Pi_i+\tilde{L}_{2,i}^{\pm}\,\tilde{R}_{1,i}^{\pm}\partial v_i\right) & \quad \text{if $\lambda_{\pm}(v_i)\ge 0$}\\
    0 & \quad \text{otherwise}
    \end{cases} 
\end{equation}
and for the velocity:
\begin{equation}
\tilde{v}_{i}^{\pm}=
    \begin{cases}
    \left(1-\frac{\Delta \tau}{\Delta x}\lambda_{\pm}(v_i)\right)\left(\tilde{L}_{1,i}^{\pm}\,\tilde{R}_{2,i}^{\pm}\partial\Pi_i+\tilde{L}_{2,i}^{\pm}\,\tilde{R}_{2,i}^{\pm}\partial v_i\right) & \quad \text{if $\lambda_{\pm}(v_i)\ge 0$}\\
    0 & \quad \text{otherwise}
    \end{cases} 
\end{equation}
where $\tilde{L}_1^{\pm}$ and $\tilde{L}_2^{\pm}$ are the elements of the left eigenvectors Eq.~(\ref{lefteigenvec}) and $\tilde{R}_1^{\pm}$ and $\tilde{R}_2^{\pm}$ are the elements of the right eigenvectors Eq.~(\ref{righteigenvec}).

The characteristic tracing is evident because we put to 0 the contributions to $\tilde{\Pi}_{i}^{\pm}$ and $\tilde{v}_{i}^{\pm}$ when the wave $\lambda_{\pm}$ is moving away from the interface that we are considering.

In this scheme we also need to enforce monotonicity of the solution. To achieve this we compute the slope $\partial W_i=\{\partial\Pi_i,\partial v_i\}$ using the SUPERBEE slope limiter:
\begin{eqnarray}
\partial W_i&=&\left[\textrm{sign}(\textbf{W}_i^{n}-\textbf{W}_{i-1}^{n})+\textrm{sign}(\textbf{W}_{i+1}^{n}-\textbf{W}_{i}^{n})\right]\times\nonumber\\
&\times&\textrm{min}\left[|\textbf{W}_i^{n}-\textbf{W}_{i-1}^{n}|,|\textbf{W}_{i+1}^{n}-\textbf{W}_{i}^{n}|,\frac{1}{2}\textrm{max}\left(|\textbf{W}_i^{n}-\textbf{W}_{i-1}^{n}|,|\textbf{W}_{i+1}^{n}-\textbf{W}_{i}^{n}|\right)\right].\nonumber\\
\end{eqnarray}

\begin{figure}[th]
\centering
\begin{tabular}{cc}
\includegraphics[width=5.5cm]{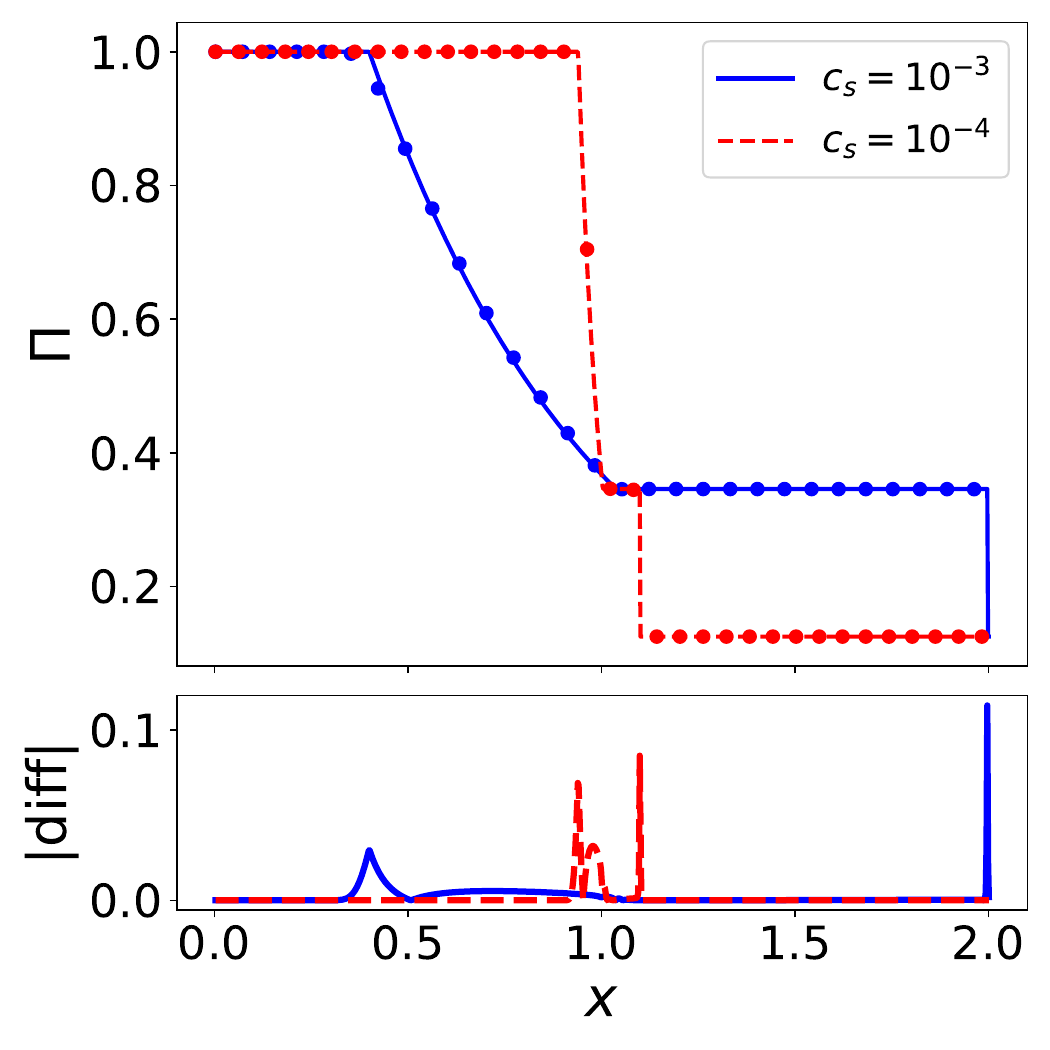} &
\includegraphics[width=5.5cm]{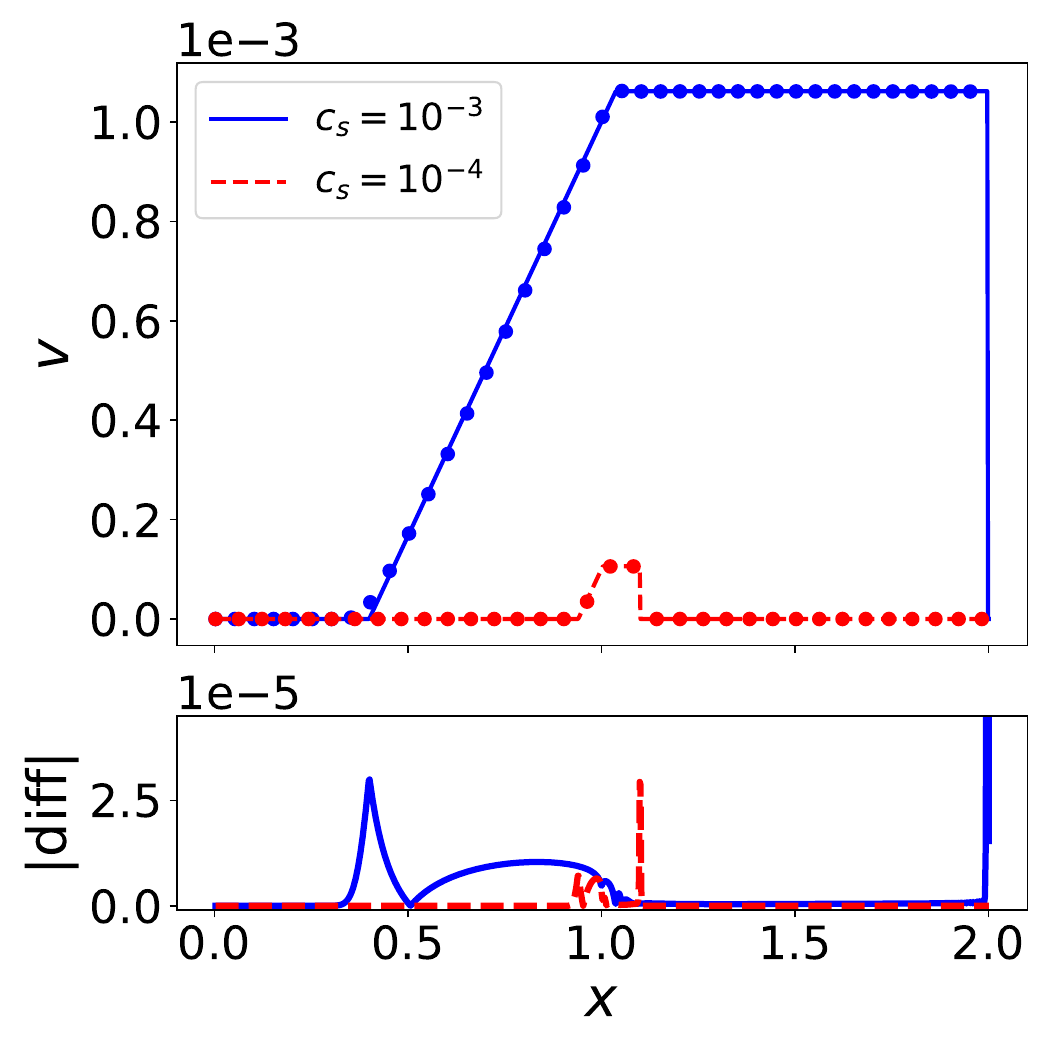} \\
\end{tabular}
\caption{PLM (circles) vs. Exact solution (lines) of Test 1 for $\Pi$ (left panel) and $v$ (right panel) at time $\tau=600$ units for $c_s=10^{-3}$ (blue solid line) and $10^{-4}$ (red dashed line) respectively. In the bottom panel the absolute difference between the numerical and exact solution.\linebreak}\label{fig8}
\end{figure}
\begin{figure}[h]
\centering
\begin{tabular}{cc}
\includegraphics[width=5.5cm]{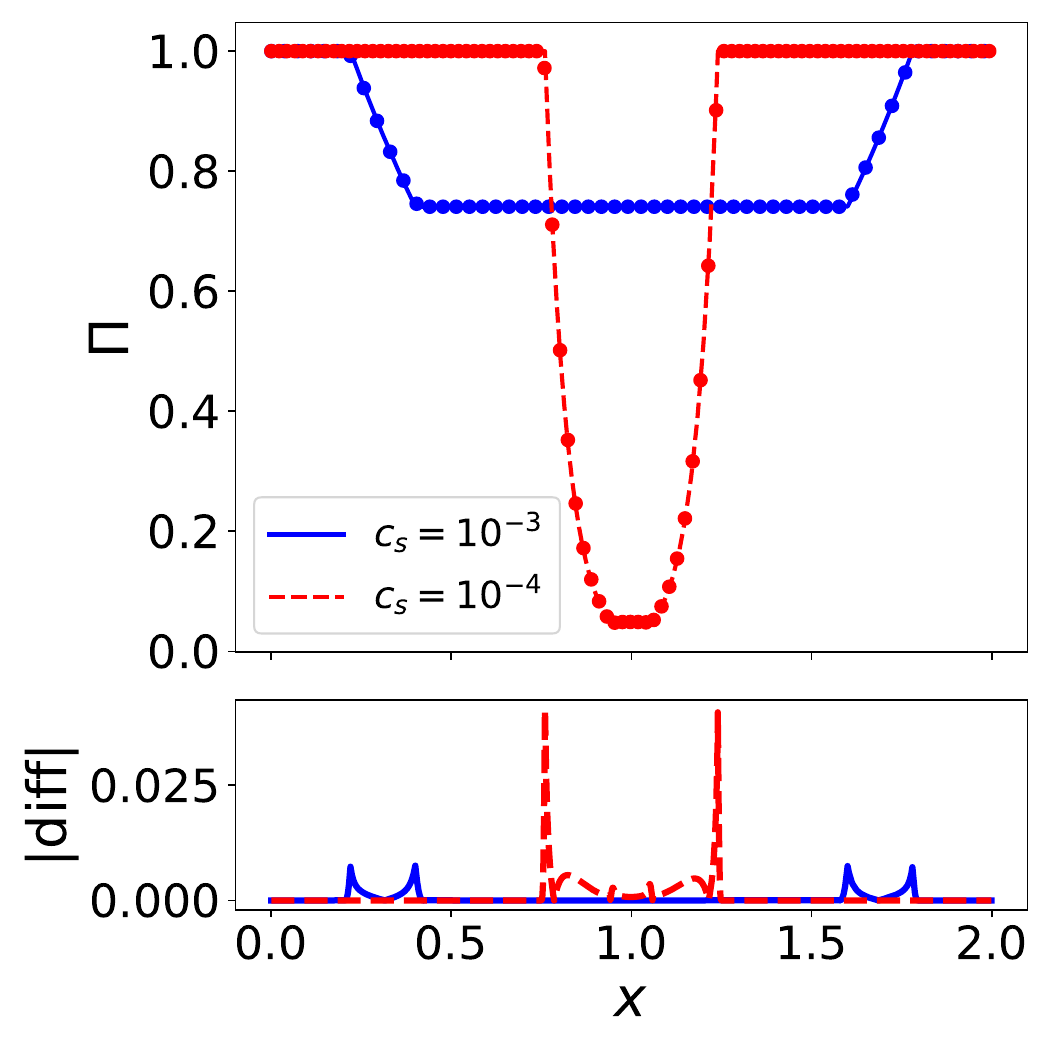}&
\includegraphics[width=5.5cm]{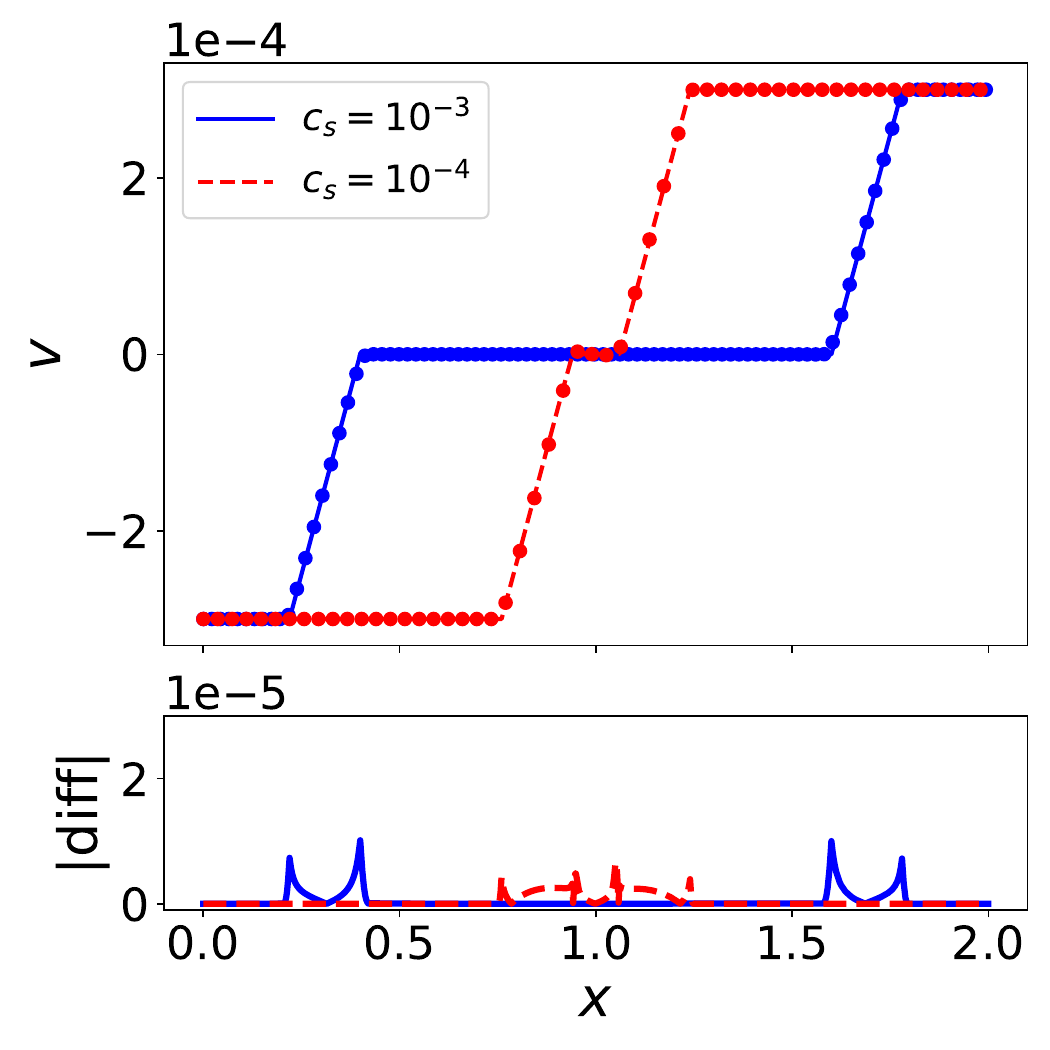} \\
\end{tabular}
\caption{As in Fig.~\ref{fig8} for Test 2.\linebreak}\label{fig9}
\end{figure}
\begin{figure}[h]
\centering
\begin{tabular}{cc}
\includegraphics[width=5.5cm]{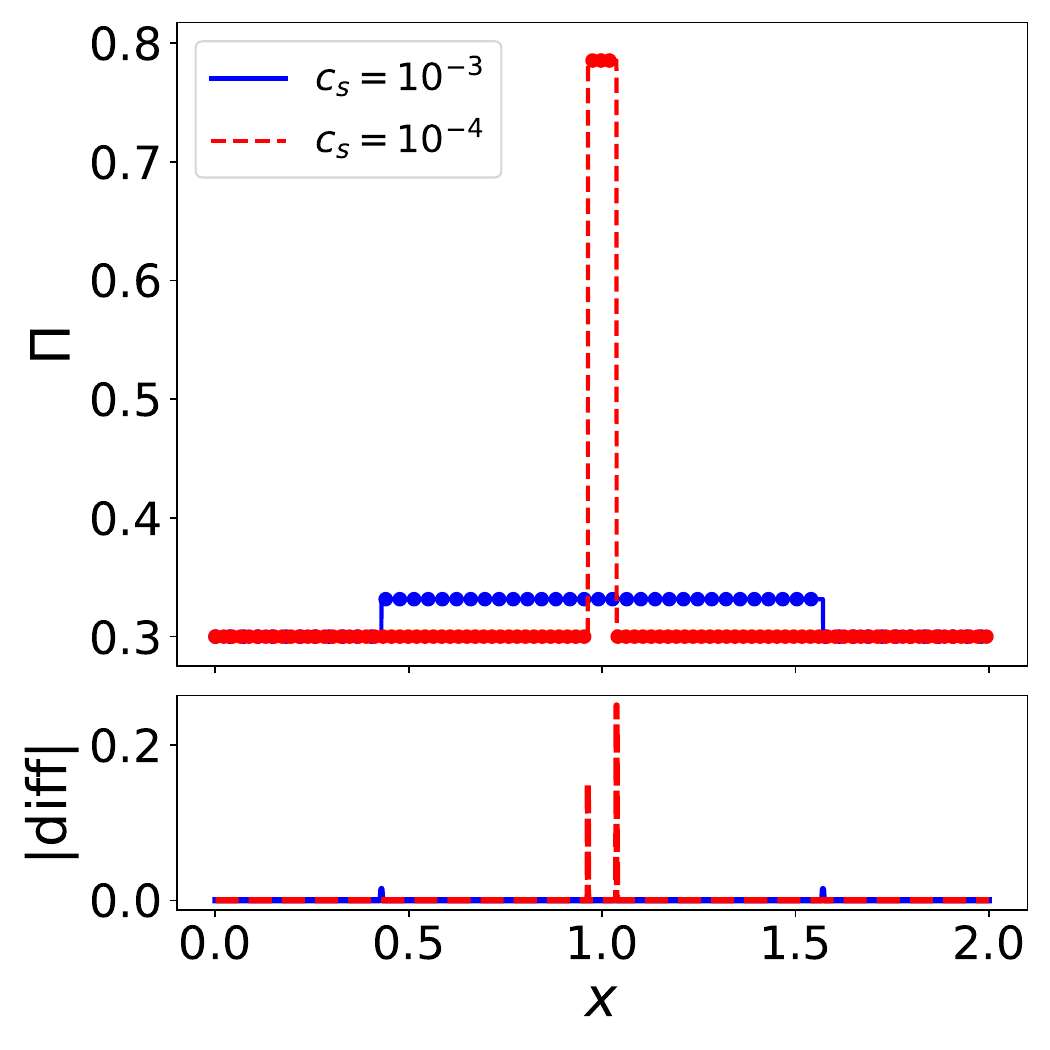}&
\includegraphics[width=5.5cm]{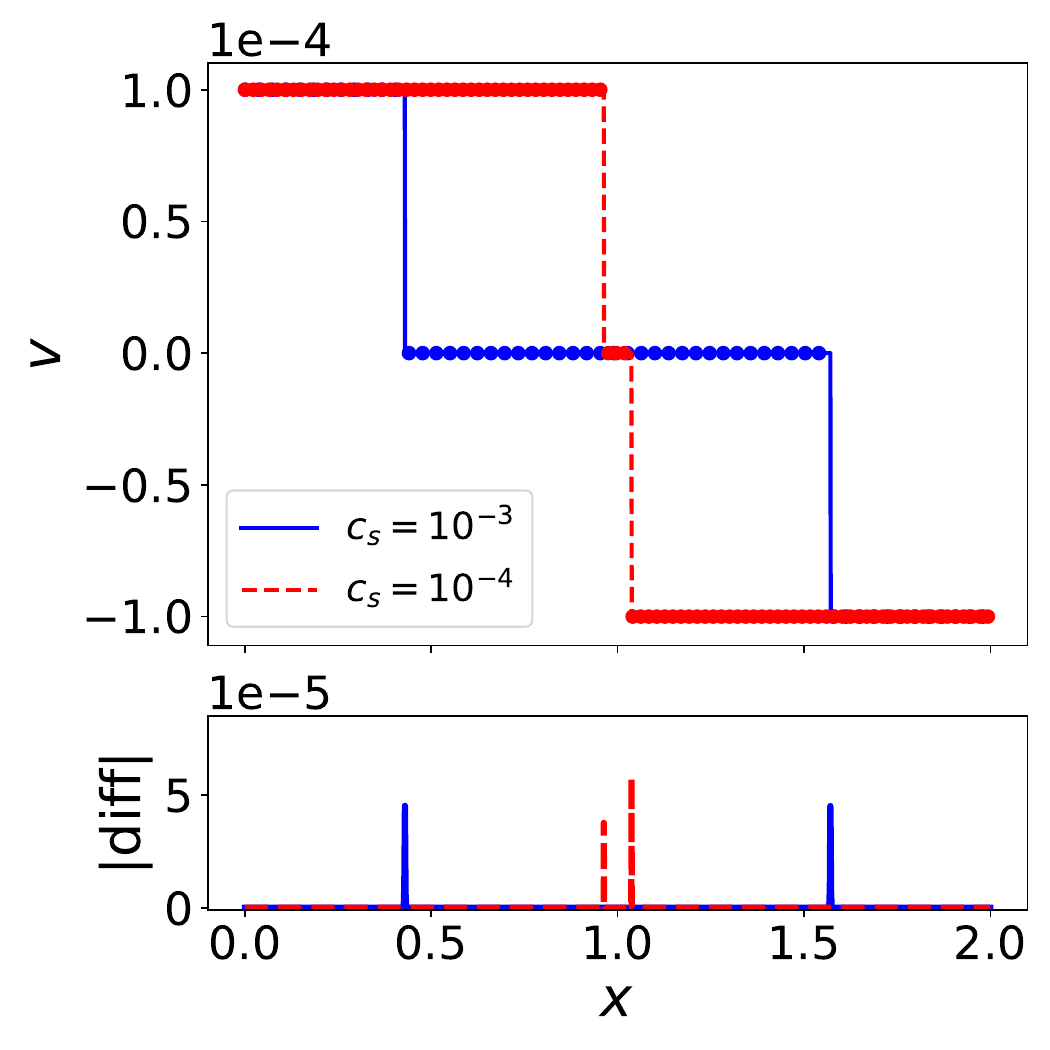} \\
\end{tabular}
\caption{As in Fig.~\ref{fig8} for Test 3.}\label{fig10}
\end{figure}

In Figs.~\ref{fig8},~\ref{fig9} and~\ref{fig10} we plot the PLM numerical solutions (dotted lines) to test cases given in Table~\ref{tab1} against the exact solutions at $\tau=600$ units for $c_s=10^{-3}$ (solid blue lines) and $c_s=10^{-4}$ (dash red lines) respectively. In the bottom panels we plot the absolute value of the difference between the numerical and exact solutions. We can see only minor differences with the respect to the results obtained with the MUSCL-Hancock method shown in Figs.~\ref{fig5},~\ref{fig6} and~\ref{fig7}. In particular, we may notice that the former is less accurate than MUSCL-Hancock in resolving rarefaction waves, while it performs better in the case of shocks. 

\begin{figure}[th]
\centering
\begin{tabular}{cc}
\includegraphics[width=5.5cm]{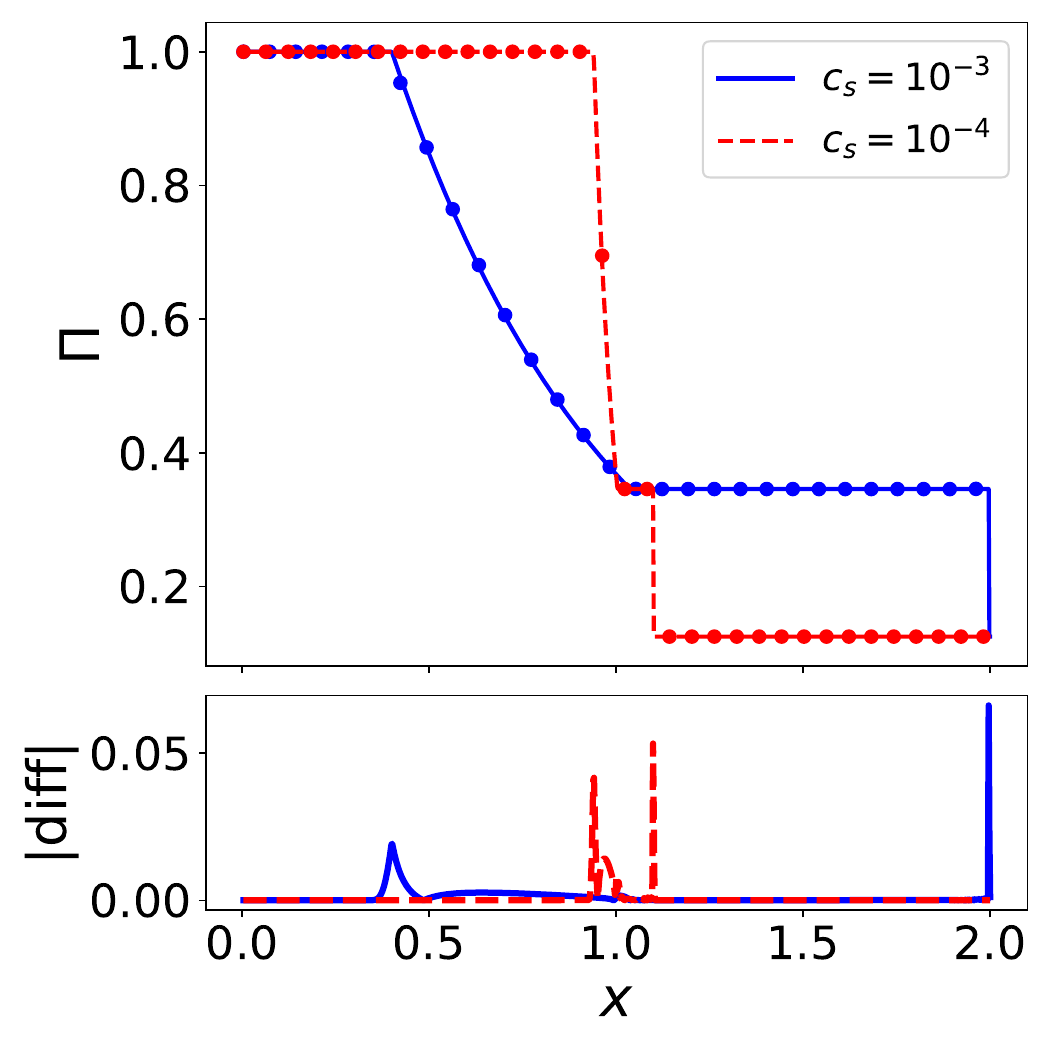} &
\includegraphics[width=5.5cm]{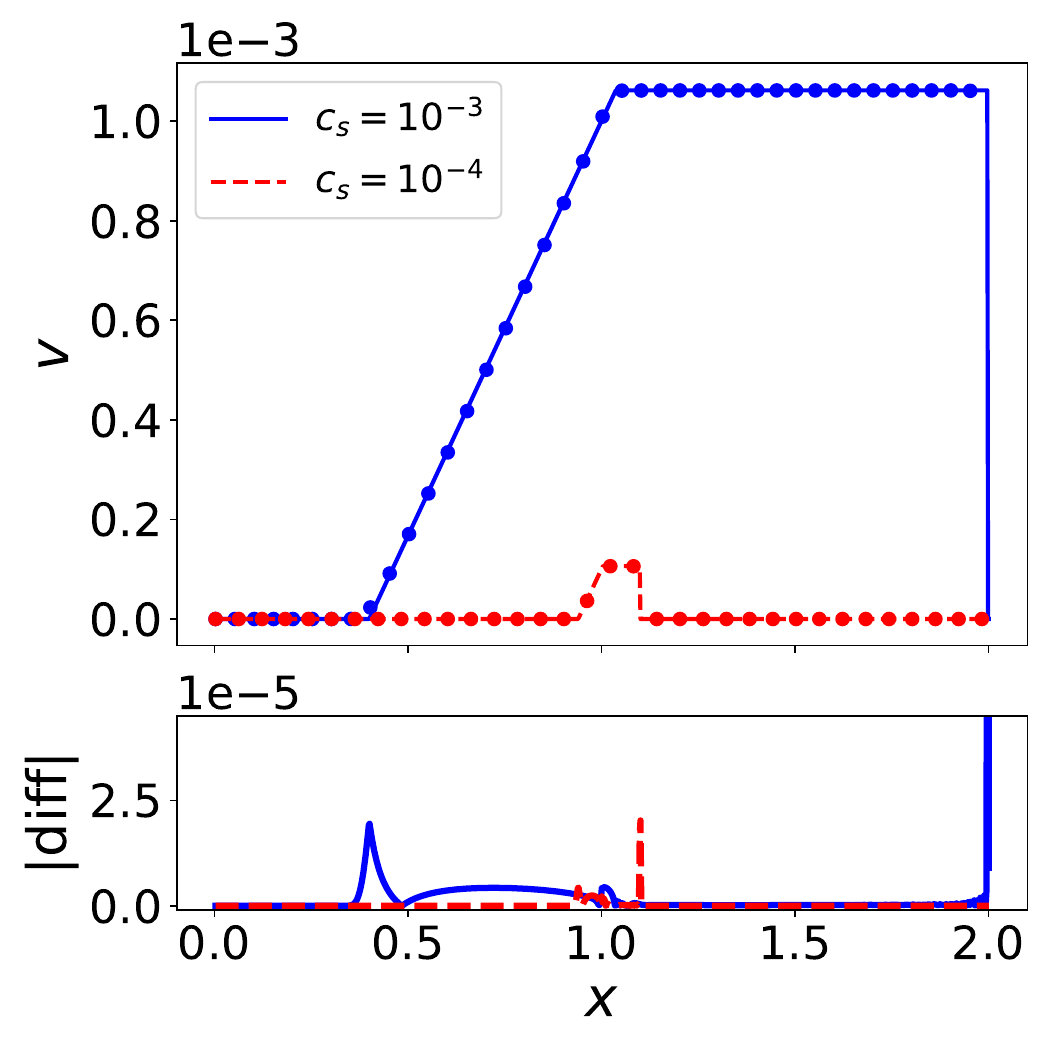} \\
\end{tabular}
\caption{PPM (circles) vs. Exact solution (lines) of Test 1 for $\Pi$ (left panel) and $v$ (right panel) at time $\tau=600$ units for $c_s=10^{-3}$ (blue solid line) and $10^{-4}$ (red dashed line) respectively. In the bottom panel the absolute difference between the numerical and exact solution.\linebreak}\label{fig11}
\end{figure}
\begin{figure}[h]
\centering
\begin{tabular}{cc}
\includegraphics[width=5.5cm]{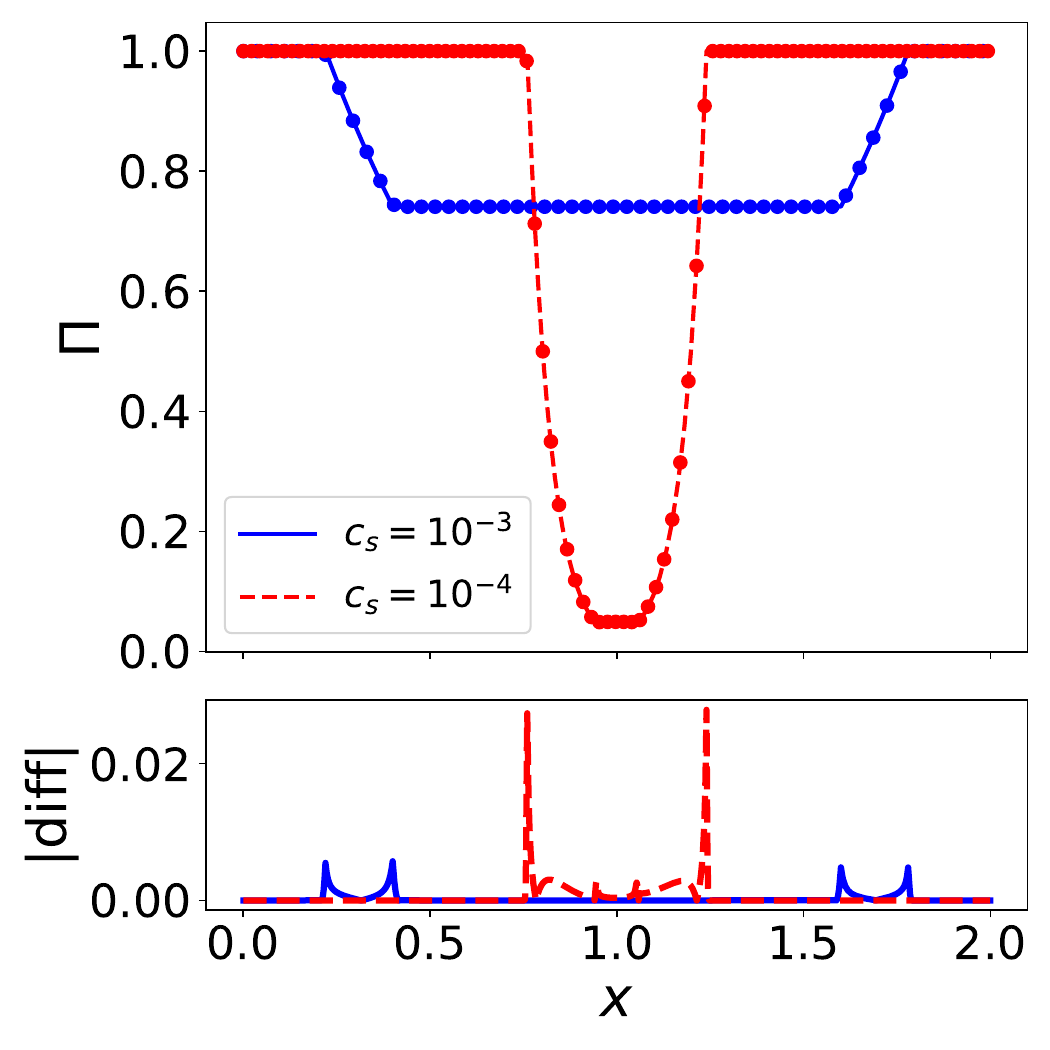}&
\includegraphics[width=5.5cm]{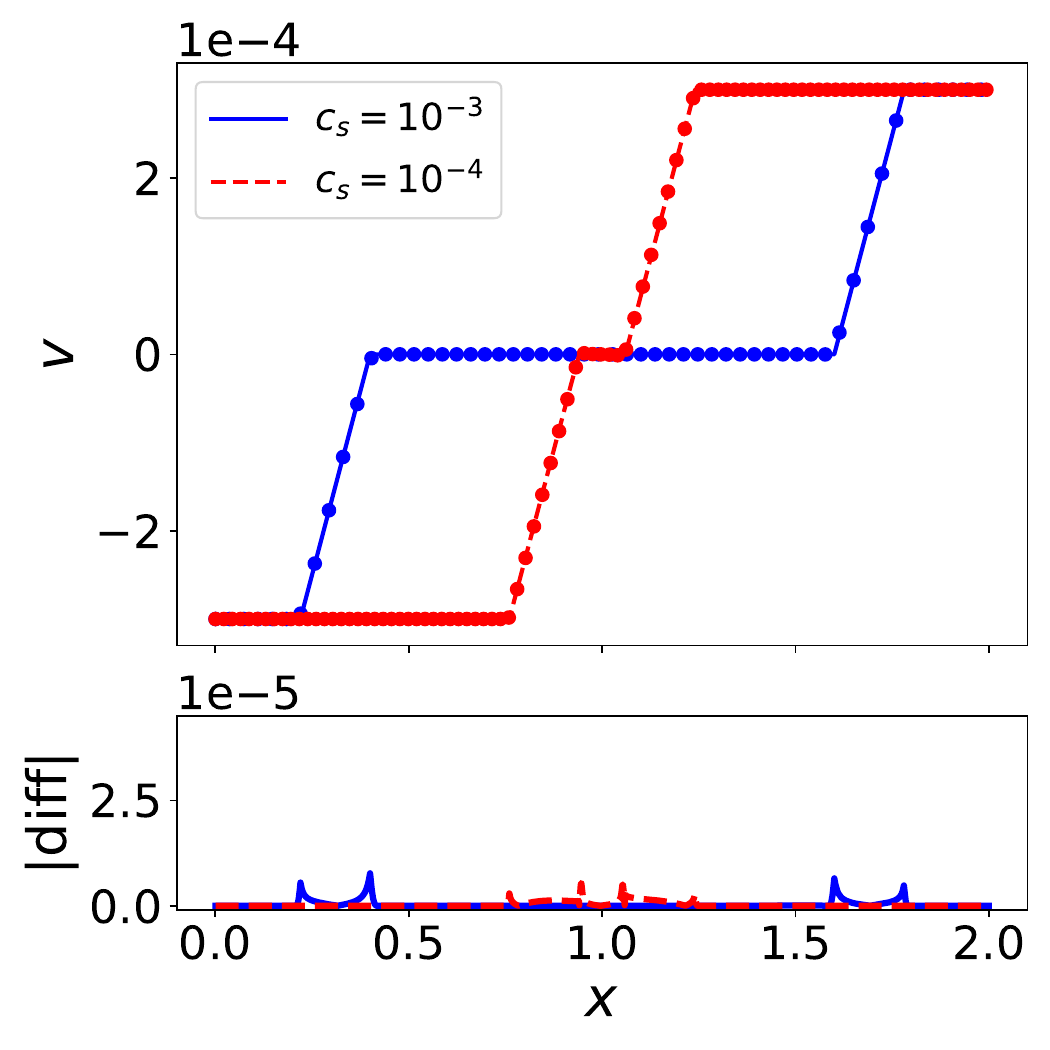} \\
\end{tabular}
\caption{As in Fig.~\ref{fig11} for Test 2.\linebreak}\label{fig12}
\end{figure}
\begin{figure}[h]
\centering
\begin{tabular}{cc}
\includegraphics[width=5.5cm]{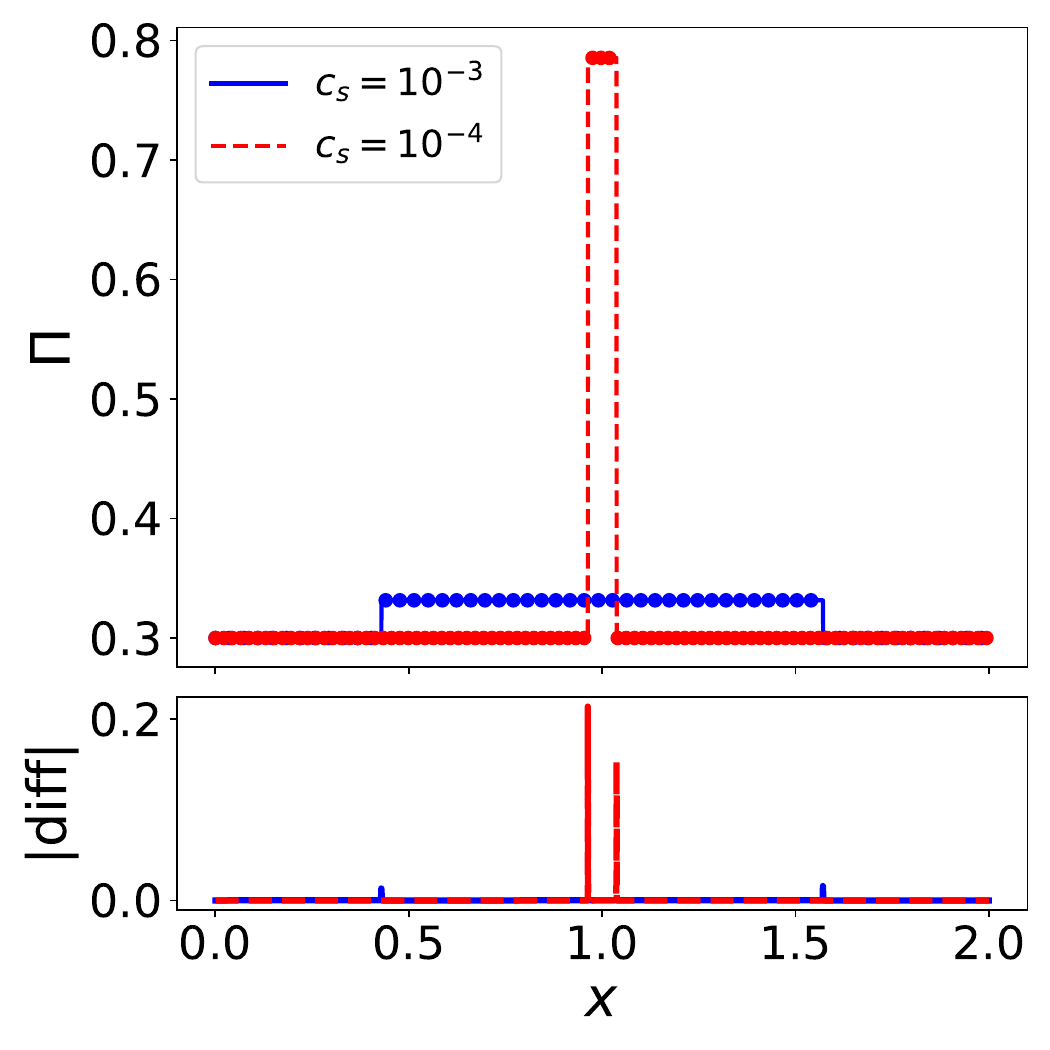}&
\includegraphics[width=5.5cm]{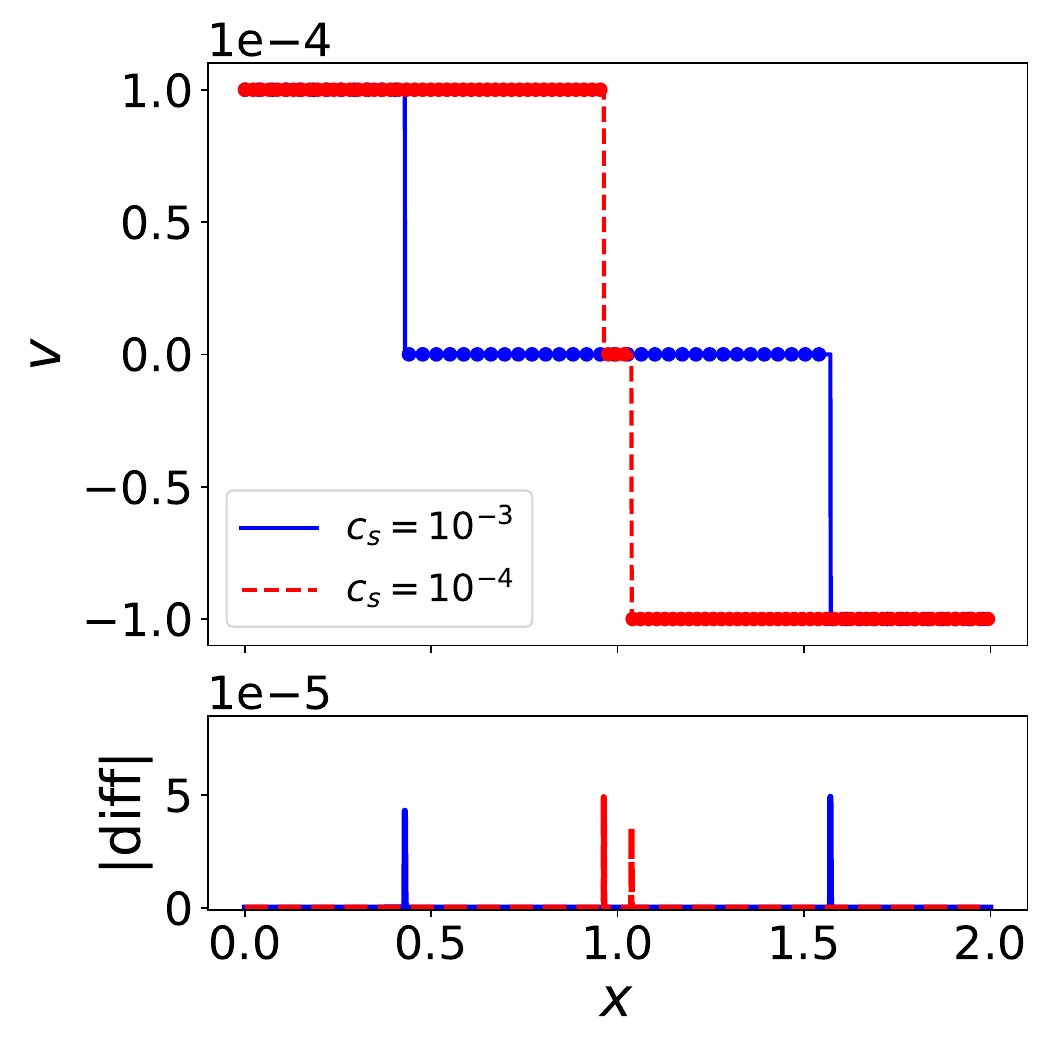} \\
\end{tabular}
\caption{As in Fig.~\ref{fig11} for Test 3.}\label{fig13}
\end{figure}

\subsection{Piecewise Parabolic Method (PPM)}\label{PPM}
Originally introduced by Colella \& Woodward \cite{ColellaWoodward1984}, this scheme achieves second-order accuracy using a parabolic reconstruction of the data with characteristic tracing. We refer again to Zingale's notes \cite{Zingale} for a clear summary of PPM in the case of non-linear advection equations. The expressions of the parabolic reconstruction of primitive variables with characteristics tracing step in the case of dark fluids are quite lengthy and here we only present the results of the numerical tests of Table~\ref{tab1} shown in Figs.~\ref{fig11},~\ref{fig12} and~\ref{fig13} (dotted lines) against the exact solutions at $\tau=600$ units for $c_s=10^{-3}$ (solid blue lines) and $c_s=10^{-4}$ (dash red lines) respectively. In the bottom panels we plot the absolute value of difference between the exact and the numerical solutions. 

\subsection{Advection Test for Higher-Order Schemes}\label{advection_test}

We consider a pure 1D advection problem and advect an initial Gaussian density profile
\begin{equation}
\Pi(x,0)\equiv\Pi_0(x)=1+e^{-\frac{(x-x_0)^2}{\sigma^2}},
\end{equation}
located at $x_0=0.5$ with $\sigma=0.1$ that moves with constant speed $v=0.1$ toward the positive values of the x-axis. The exact solution to the advection equation is simply $\Pi(x,\tau)=\Pi_0(x-v\cdot\tau)$.

\begin{figure}[h]
\centering
\begin{tabular}{cc}
\includegraphics[width=0.5\textwidth]{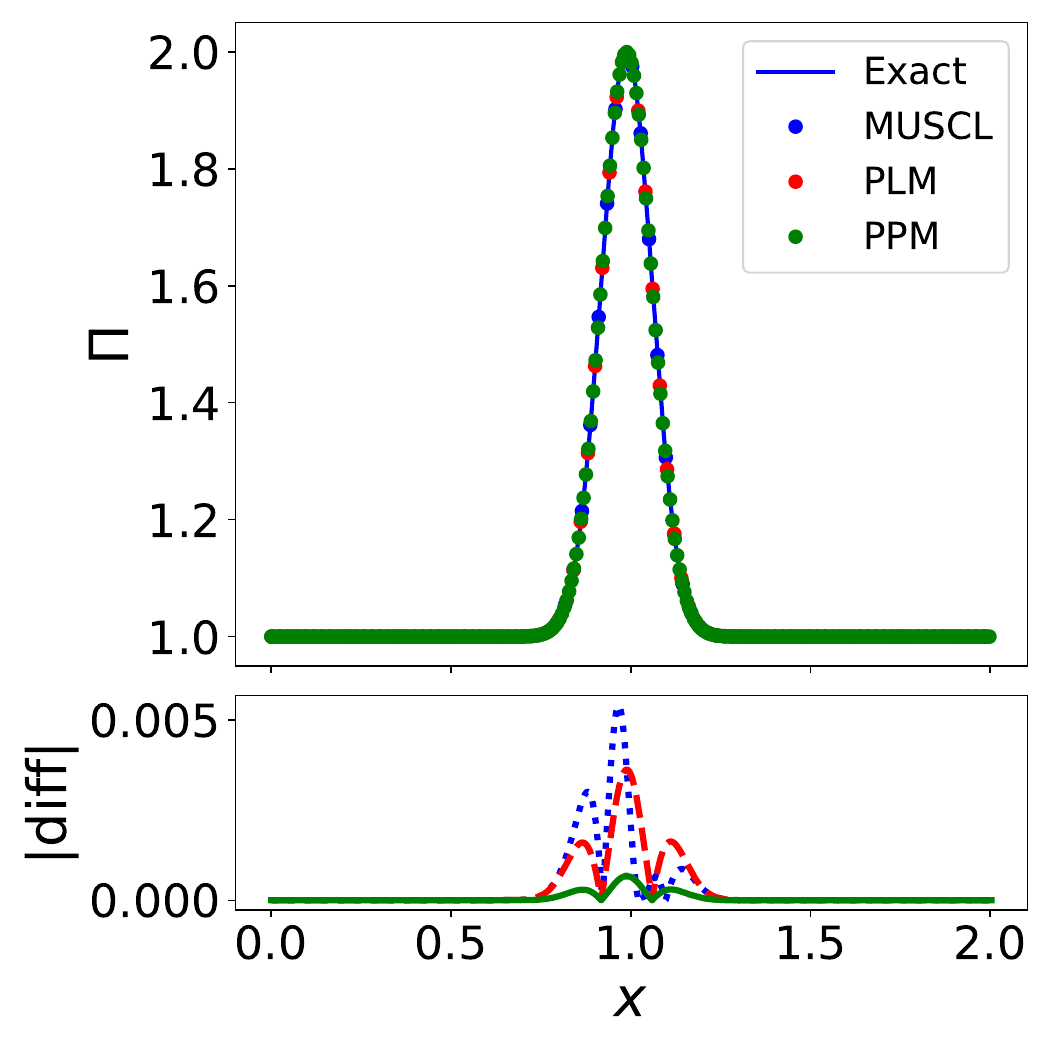}
\end{tabular}
\caption{Advection of a Gaussian density profile (located at $x_0=0.5$ at the initial time) moving at constant speed $v=0.1$ at time $\tau=5$ units for $c_s=10^{-3}$. Shown in the plot are the exact solution (blue solid line) and the numerical solutions obtained using the MUSCL-Hancock (blue points), PLM (red points) and PPM (green points) schemes respectively. In the bottom panel we plot the absolute difference between the exact and the numerical solutions.}\label{fig_advection}
\end{figure}

We solve numerically the 1D advection equation for the initial density profile $\Pi_0(x)$ in the case of $c_s=10^{-3}$ with periodic boundary condition using the MUSCL-Hancock method, the PLM and PPM schemes previously discussed. The results are shown in Fig.~\ref{fig_advection}, where we plot the exact solution at time $\tau=5$ (blue solid line) and that obtained using MUSCL-Hancock (blue points), PLM (red points) and PPM (green points). The bottom panel shows the absolute difference with respect to the exact solution. We can see the improvement in accuracy of the PPM scheme compared to the PLM and MUSCL-Hancock methods.

\section{Conclusions}\label{conclu}
A dynamical form of dark energy is expected to lead to dark energy inhomogeneities whose properties are characterized by an effective speed of sound parameter. Cosmological observations of the cosmic microwave background as well as large-scale structure have so far left the dark energy speed of sound unconstrained. However, ongoing and future galaxy surveys may have the potential to constrain this parameter, especially in the case of dark energy models with small sound speed values for which dark energy perturbations can become significant at small scales and late time. The detection of the effects of such inhomogeneities can provide smoking gun evidence against the cosmological constant paradigm, thus unveiling the dynamical nature of this exotic component. Nevertheless, this requires the availability of theoretical predictions capable of following the gravitational collapse of the coupled system of dark matter and dark energy throughout the non-linear regime of cosmic structure formation.

The gravitational collapse of dark matter density fluctuations can be followed using the standard N-body method. However, this is not the case for a dark energy fluid characterized by a negative equation of state and a non-vanishing sound speed. This is because in a given cosmological volume the mass of the particles tracing the clustering dark energy field is not conserved, rather it varies in space and time in a way that cannot be determined by the N-body equations of motion in a closed form. In contrast, it is possible to directly solve the Euler equations describing the evolution of the dark energy fluid using hydrodynamical methods. Matter and dark energy perturbations are mutually coupled gravitationally via the Poisson equation for the gravitational potential.

Finite volume conservative methods have been developed in a vast literature to numerically solve hydrodynamic equations. These are implemented in established cosmological simulation codes that simulate the gravitational collapse of baryonic fluids. Hence, the use of such methods to solve the dark energy fluid equations can facilitate their implementation in simulation codes, thus allowing to perform full cosmological simulations of clustering dark energy models. One of the most widely adopted conservative numerical schemes is the Godunov method. This relies on the solution of the Riemann Problem of the advection part of the hydrodynamic equations considered. As such, the use of such a method to solve the dynamics of dark energy fluctuation in a cosmological simulation code requires the derivation of exact and approximate solutions that are specific to the form of the dark energy fluid equations.

In this work we have presented a reformulation of the Euler equations describing a dark energy fluid in a quasi-conservative form of hyperbolic partial differential equations with source terms. We have derived the exact solution of the Riemann Problem associated with the advection part of these equations in the non-relativistic limit with a small non-vanishing speed of sound. Physically, this corresponds to finding the wave structure of dark fluid perturbations propagating in a non-expanding background in the absence of gravitational interactions.

We have constructed a number of approximate solvers of the RP specific to the dark energy fluid equations and tested their accuracy against exact solutions of standard test cases, such as the Sod test, the propagation of two rarefaction waves and two shock waves. The use of such approximate solvers is necessary to significantly speed up the computation of the numerical solution of the Euler equations with the Godunov method. We have presented the equations describing the original formulation of the Godunov method that is first-order accurate in space and time. We have also derived the formula specific to the Euler equations of the dark fluid considered here for higher-order schemes based on the Godunov method, such as the MUSCL-Hancock method, the Piecewise Linear Method and the Piecewise Parabolic Method. We have tested the accuracy of these methods against exact solutions of standard test cases considered for the RP as well as an advection test.

In a follow-up work we will present the full cosmological simulation code, where we account for the cosmological source terms as well as the gravitational interaction and test the numerical solution in 3D.

To conclude, this work lays down the first stepping stone to constrain the speed of sound of Dark Energy perturbations using large-scale-structure data and opens the way to learning more about the Dark Energy phenomenon.

\begin{acknowledgments}
We thank Eiichiro Komatsu, Emiliano Sefusatti, Volker Springel and Romain Teyssier for useful discussions, and the anonymous referee for helpful comments which lead to a clarified presentation. The research leading to these results has received funding from the European Research Council under the European Community's Seventh Framework Programme (FP7/2007-2013 Grant Agreement no. 279954). 
LB and FS acknowledge support from the Starting Grant (ERC-2015-STG 678652) ``GrInflaGal'' of the European Research Council.
\end{acknowledgments}
\appendix
\section{Elementary Wave Solutions in 1D}\label{app_rp1d}
\subsection{Rarefaction Waves}
We can use the Riemann invariants Eq.~(\ref{RI}) to relate a known state to the left (right) of a rarefaction wave to that on the right (left). These relations can be written in terms of primitive variables as:
\begin{eqnarray}
    \frac{d\Pi}{\Pi}&=&\frac{\left(1+\frac{c_s^2}{c^2}\right)dv}{c_s\sqrt{1-\frac{v^2}{c^2}}-\frac{c_s^2}{c^2}v} \quad (\lambda=\lambda_+)\\
    \frac{d\Pi}{\Pi}&=&\frac{\left(1+\frac{c_s^2}{c^2}\right)dv}{-c_s\sqrt{1-\frac{v^2}{c^2}}-\frac{c_s^2}{c^2}v} \quad (\lambda=\lambda_-)   
\end{eqnarray}
and can be integrated analytically to obtain:
\begin{eqnarray}
    \ln{\Pi}&-&\frac{c}{c_s}\arcsin\left(\frac{v}{c}\right)+\ln\left(\sqrt{1-\frac{v^2}{c^2}}-\frac{v c_s}{c^2}\right)=const. \quad (\lambda=\lambda_+)\nonumber \\\label{rightw}\\
    \ln{\Pi}&+&\frac{c}{c_s}\arcsin\left(\frac{v}{c}\right)+\ln\left(\sqrt{1-\frac{v^2}{c^2}}+\frac{v c_s}{c^2}\right)=const.  \quad (\lambda=\lambda_-)\nonumber \\ \label{leftw} 
\end{eqnarray}
which express the conservation of these quantities across the waves associated to the characteristic fields $\lambda_+$ and $\lambda_-$ respectively.

\subsubsection{Left Rarefaction}
Let us consider a left rarefaction wave associated with $\lambda_-$-field. Using the Riemann invariant Eq.~(\ref{leftw}) we can relate the known state $\textbf{U}_{\rm L}$ to the unknown state in the star region $\textbf{U}^*$ shown in Fig.~\ref{fig1}. It is useful to introduce the auxiliary function
\begin{equation}\label{glr_def}
g_{\rm Lr}(v)\equiv\frac{c}{c_s}\arcsin\left(\frac{v}{c}\right)+\ln\left(\sqrt{1-\frac{v^2}{c^2}}+\frac{v c_s}{c^2}\right),
\end{equation}
hence from Eq.~(\ref{leftw}) we obtain
\begin{equation}
g_{\rm Lr}(v^*)=\ln{\frac{\Pi_{\rm L}}{\Pi^*}}+g_{\rm Lr}(v_{\rm L}).\label{glr}
\end{equation}
This implicit relation can be inverted to obtain 
\begin{equation}
v^*=f_{\rm Lr}(\Pi^*,\textbf{W}_{\rm L}),\label{flr}
\end{equation}
where we have denoted with $f_{\rm Lr}$ the inverse function $g_{\rm Lr}^{-1}$.

The rarefaction wave is enclosed in a region delimited by the $\textit{head}$ of the wave moving with speed $s_{\rm HLr}=\lambda_-(v_{\rm L})$ and the $\textit{tail}$ with speed $s_{\rm TLr}=\lambda_-(v^*)$. The evolution of this region, also dubbed $\textit{rarefaction fan}$, can be obtained by integrating the equation $dx/d\tau=\lambda_-(v)$ along the characteristic with initial condition $x(0)=x_0$ (the location of the discontinuity at initial time). This gives
\begin{equation}
\frac{x-x_0}{\tau}=\lambda_-(v_{\rm L fan}),\label{leftvfan}
\end{equation}
which can be inverted to obtain $v_{\rm L fan}(x)$ at any given time. $\Pi_{\rm L fan}(x)$ inside the fan is then obtained using the generalized Riemann invariant Eq.~(\ref{leftw}):
\begin{equation}
\ln{\Pi_{\rm L fan}}(x)=\ln{\Pi_{\rm L}}+g_{\rm Lr}(v_{\rm L})-g_{\rm Lr}(v_{\rm L fan}(x)).\label{leftdfan}
\end{equation}

\subsubsection{Right Rarefaction}
Let us now consider a right rarefaction wave associated with the characteristic field $\lambda_+$. Using the Riemann invariant Eq.~(\ref{rightw}) we can relate the known state $\textbf{U}_{\rm R}$ to the right of the wave to the unknown state in the star region $\textbf{U}^*$ shown in Fig.~\ref{fig1}. Let us introduce the auxiliary function
\begin{equation}\label{grr_def}
g_{\rm Rr}(v)\equiv-\frac{c}{c_s}\arcsin\left(\frac{v}{c}\right)+\ln\left(\sqrt{1-\frac{v^2}{c^2}}-\frac{v c_s}{c^2}\right),
\end{equation}
from Eq.~(\ref{leftw}) we obtain
\begin{equation}
g_{\rm Rr}(v^*)=\ln{\frac{\Pi_{\rm R}}{\Pi^*}}+g_{\rm Rr}(v_{\rm R}).\label{grr}
\end{equation}
This implicit relation can be inverted to obtain 
\begin{equation}
v^*=f_{\rm Rr}(\Pi^*,\textbf{W}_{\rm R}),\label{frr}
\end{equation}
where we have denoted with $f_{\rm Rr}$ the inverse function $g_{\rm Rr}^{-1}$.

The right rarefaction fan is delimited by the head of the wave moving with speed $s_{\rm HRr}=\lambda_+(v_{\rm R})$ and the tail with speed $s_{\rm TRr}=\lambda_+(v^*)$. The evolution of the fan region is given by solving $dx/d\tau=\lambda_+(v)$ with $x(0)=x_0$ (the location of the discontinuity at initial time):
\begin{equation}
\frac{x-x_0}{\tau}=\lambda_+(v_{\rm R fan}),\label{rightvfan}
\end{equation}
this can be inverted to infer $v_{\rm R fan}(x)$. Then, using the Riemann invariant Eq.~(\ref{rightw}) we obtain $\Pi_{\rm R fan}(x)$ inside the fan:
\begin{equation}
\ln{\Pi_{\rm R fan}(x)}=\ln{\Pi_{\rm R}}+g_{Rr}(v_{\rm R})-g_{\rm Lr}(v_{\rm R fan}(x)).\label{rightdfan}
\end{equation}

\subsection{Shock Waves}
We can use the RH conditions in Eq.~(\ref{RKHU}) to relate a known state to the left (right) of the shock to that unknown to the right (left).

\subsubsection{Left Shock}
Let us consider a shock moving from right to left separating the known initial state $\textbf{U}_L$ from the unknown state $\textbf{U}^*$ in the star region represented in Fig.~\ref{fig1}. Using Eq.~(\ref{RKHU}) we have a system of algebraic equations which reads as
\begin{eqnarray}
\left(1+\frac{c_s^2}{c^2}\right)(\Pi_{\rm L} v_{\rm L}-\Pi^* v^*)&=&s_{\rm Ls}(\Pi_{\rm L}-\Pi^*)\label{RH1l}\\
\frac{c_s^2}{1+\frac{c_s^2}{c^2}}(\Pi_{\rm L}-\Pi^*)+(\Pi_{\rm L} v_{\rm L}^2-\Pi^* {v^*}^2)&=&s_{\rm Ls}(\Pi_{\rm L} v_{\rm L}-\Pi^* v^*).\label{RH2l}
\end{eqnarray}
This is an algebraic system of two equations and three unknowns ($\Pi^*,v^*,s_{\rm Ls}$). We can solve for $s_{\rm Ls}$ and $v^*$ to find
\begin{eqnarray}
\frac{s_{\rm Ls}}{1+\frac{c_s^2}{c^2}}&=&
\frac{\Pi_{\rm L} v_{\rm L}}{\Pi_{\rm L}+\frac{c_s^2}{c^2}\Pi^*}-
\sqrt{ \frac{\Pi_{\rm L}^2 v_{\rm L}^2}{(\Pi_{\rm L}+\frac{c_s^2}{c^2}\Pi^*)^2}-
\frac{\Pi_{\rm L} v_{\rm L}^2}{\Pi_{\rm L}+\frac{c_s^2}{c^2}\Pi^*}+
\frac{\Pi^* c_s^2}{\left(\Pi_{\rm L}+\frac{c_s^2}{c^2}\Pi^*\right)\left(1+\frac{c_s^2}{c^2}\right)}
}\nonumber\\
\label{leftvshock} \\
v^*&\equiv&f_{\rm Ls}(\Pi^*,\textbf{W}_{\rm L})=\frac{\Pi_{\rm L}}{\Pi^*}v_{\rm L}-\frac{s_{\rm Ls}}{1+\frac{c_s^2}{c^2}}\,\frac{\Pi_{\rm L}-\Pi^*}{\Pi^*},\label{left_shock_rel}
\end{eqnarray}
where $s_{\rm Ls}$ is the negative root of the quadratic equation obtained from Eqs.~(\ref{RH1l}) and~(\ref{RH2l}), consistently with the fact that the left shock moves from the right to the left.

\subsubsection{Right shock}
Let us consider a shock moving from the left to the right separating the known initial state $\textbf{U}_R$ from the unknown state $\textbf{U}^*$ in the star region represented in Fig.~\ref{fig1}. Using the Rankine-Hugoniot conditions we obtain the system of algebraic equations
\begin{eqnarray}
\left(1+\frac{c_s^2}{c^2}\right)(\Pi^* v^*-\Pi_{\rm R} v_{\rm R})&=&s_{\rm Rs}(\Pi^*-\Pi_{\rm R})\label{RH1r}\\
\frac{c_s^2}{1+\frac{c_s^2}{c^2}}(\Pi^*-\Pi_{\rm R})+(\Pi^* {v^*}^2-\Pi_{\rm R} v_{\rm R}^2)&=&s_{\rm Rs}(\Pi^* v^*-\Pi_{\rm R} v_{\rm R}).\label{RH2r}
\end{eqnarray}
This is an algebraic system of two equations and three unknowns ($\Pi^*,v^*,s_{\rm Rs}$), solving for $s_{\rm Rs}$ and $v^*$ we find
\begin{eqnarray}
\frac{s_{\rm Rs}}{1+\frac{c_s^2}{c^2}}&=&
\frac{\Pi_{\rm R} v_{\rm R}}{\Pi_{\rm R}+\frac{c_s^2}{c^2}\Pi^*}+
\sqrt{
\frac{\Pi_{\rm R}^2 v_{\rm R}^2}{(\Pi_{\rm R}+\frac{c_s^2}{c^2}\Pi^*)^2}-
\frac{\Pi_{\rm R} v_{\rm R}^2}{\Pi_{\rm R}+\frac{c_s^2}{c^2}\Pi^*}+
\frac{\Pi^* c_s^2}{\left(\Pi_{\rm R}+\frac{c_s^2}{c^2}\Pi^*\right)\left(1+\frac{c_s^2}{c^2}\right)}
}\nonumber\\
\label{rightvshock}\\
v^*&\equiv&f_{\rm Rs}(\Pi^*,\textbf{W}_{\rm R})=\frac{\Pi_{\rm R}}{\Pi^*}v_{\rm R}+\frac{s_{\rm Rs}}{1+\frac{c_s^2}{c^2}}\,\frac{\Pi^*-\Pi_{\rm R}}{\Pi^*},\label{right_shock_rel}
\end{eqnarray}
where $s_{\rm Rs}$ is the positive root of the quadratic equation obtained from Eqs.~(\ref{RH1r}) and~(\ref{RH2r}), consistently with the fact that the right shock moves from the left to the right. 
\section{Elementary Wave Solutions in 3D}\label{app_rp3d}
\subsection{Rarefaction Waves}
Defining:
\begin{equation}
f_{1,4}\left(v_x\right)=1\pm\frac{v_x c_s}{c^2\sqrt{1-\frac{v_x^2}{c^2}}}
\end{equation}
the Riemann Invariants for the $\lambda_{1,4}$ eigenvalue are given by:
\begin{equation}
\frac{d \Pi}{1+\frac{c_s^2}{c^2}}=\frac{d (\Pi v_x)}{\lambda_1}=\frac{d (\Pi v_y)}{v_y\,f_{1,4}\left(v_x\right)}=\frac{d(\Pi v_z)}{v_z\,f_{1,4}\left(v_x\right)}.
\end{equation}
From the first equality we see that the Riemann invariant for $\Pi$ and $v_x$ remain the same as the 1D case. To find the values of $v_y$ across the wave we can solve:
\begin{equation}
\frac{d \Pi}{1+\frac{c_s^2}{c^2}}f_{1,4}\left(v_x\right)=\frac{d (\Pi v_y)}{v_y}
\end{equation}
By eliminating $\Pi$ using the first equality we can get to an equation that relates $v_y$ to $v_x$:
\begin{equation}
\frac{
- v_x \pm c_s\sqrt{1-\frac{v_x^2}{c^2}}
}{
c^2-v_x^2\pm c_s v_x\sqrt{1-\frac{v_x^2}{c^2}}
}
dv_x=
\frac{
dv_y
}{
v_y
}
\end{equation}
which can be integrated to give:
\begin{equation}
\frac{\frac{v_y}{c}}{\pm \frac{c_s v_x}{c^2}+\sqrt{1-\frac{v_x^2}{c^2}}}=const.
\end{equation}

Similarly, for the third component of the velocity $v_z$:
\begin{equation}
\frac{\frac{v_z}{c}}{\pm \frac{c_s v_x}{c^2} +\sqrt{1-\frac{v_x^2}{c^2}}}=const.
\end{equation}
We can then find the values of the transverse velocities in the star region as:
\begin{align}
v^*_{y;L,R}&=v_{y;L,R}\frac{\pm \frac{c_s u^*}{c^2}+\sqrt{1-\frac{{v_x^*}^2}{c^2}}}{\pm \frac{c_s v_{x;L,R}}{c^2}+\sqrt{1-\frac{v_{x;L,R}^2}{c^2}}}\\
v^*_{z;L,R}&=v_{z;L,R}\frac{\pm \frac{c_s v_x^*}{c^2}+\sqrt{1-\frac{{v_x^*}^2}{c^2}}}{\pm \frac{c_s v_{x;L,R}}{c^2}+\sqrt{1-\frac{{v_x^*}^2}{c^2}}}
\end{align}
Similarly, to find the solution inside the rarefaction fan we make use of the Riemann Invariants and compute:
\begin{align}
v^{\rm fan}_{y;L,R}&=v_{y;L,R}\frac{\pm \frac{c_s v^{\rm fan}_{x;L,R}}{c^2}+\sqrt{1-\frac{{v^{\rm fan}_{x;L,R}}^2}{c^2}}}{\pm \frac{c_s v_{x;L,R}}{c^2}+\sqrt{1-\frac{v_{x;L,R}^2}{c^2}}}\\
v^{\rm fan}_{z;L,R}&=v_{z;L,R}\frac{\pm \frac{c_s v^{\rm fan}_{x;L,R}}{c^2}+\sqrt{1-\frac{{v^{\rm fan}_{x;L,R}}^2}{c^2}}}{\pm \frac{c_s v_{x;L,R}}{c^2}+\sqrt{1-\frac{v_{x;L,R}^2}{c^2}}}
\end{align}

\subsection{Shock Waves}
The Rankine-Hugoniot conditions in this case read:
\begin{align}
\left(1+\frac{c_s^2}{c^2}\right)(\Pi_{L,R} v_{x;L,R}-\Pi^* v_x^*)&=s_{L,R;s} (\Pi_{L,R} - \Pi^*)\label{rh1} \\
\Pi_{L,R} v_{x;L,R}^2-\Pi^* {v_x^*}^2\pm\frac{c_s^2}{1+\frac{c_s^2}{c^2}}(\Pi_{L,R} - \Pi^*)&=s_{L,R;s}(\Pi_{L,R} v_{x;L,R} - \Pi^* v_x^*)\label{rh2} \\
\Pi_{L,R}\, v_{x;L,R}\, v_{y;L,R} - \Pi^* v_x^* v_y^*&=s_{L,R;s}(\Pi_{L,R} \,v_{y;L,R} - \Pi^* v_y^*)\label{rh3}\\
\Pi_{L,R}\, v_{x;L,R}\, v_{z;L,R} - \Pi^* v_x^* v_z^*&=s_{L,R;s}(\Pi_{L,R} \,v_{z;L,R} - \Pi^* v_z^*)\label{rh4}
\end{align}
Again we see that the solution for $\Pi$ and $v_x$ remains identical to the 1D case, while we can easily solve Equations~\ref{rh3} and \ref{rh4} to find:
\begin{eqnarray}
v^*_{y;L,R}=\Pi_{L,R}\, v_{y;L,R} \frac{s_{L,R;s}-v_{x;L,R}}{\Pi^*(s_{L,R;s}-v_x^*)}\label{v*Ls}\\
v^*_{z;L,R}=\Pi_{L,R}\, v_{z;L,R} \frac{s_{L,R;s}-v_{x;L,R}}{\Pi^*(s_{L,R;s}-v_x^*)}\label{w*Ls}
\end{eqnarray}

\bibliographystyle{JHEP}
\bibliography{edecs_paperI}
\end{document}